\documentclass[runningheads]{article}

\usepackage{fullpage}
\usepackage[utf8]{inputenc}

\usepackage[hyphens]{url}
\usepackage{color}
\usepackage{xcolor}
\usepackage{graphicx}
\usepackage{amsmath}
\usepackage{amsfonts}
\usepackage{tabularx}
\usepackage{booktabs}
\usepackage{framed}
\usepackage{multirow}
\usepackage[inline,shortlabels]{enumitem}
\usepackage[capitalise,noabbrev,nameinlink]{cleveref}
\usepackage{subfig}
\usepackage{tikz}
\usepackage[demo]{tikzpeople}
\usepackage{balance}
\usepackage{authblk}

\renewcommand{\paragraph}[1]{\medskip\noindent\textbf{#1}}

\newtheorem{definition}{Definition}

\renewcommand{\paragraph}[1]{\medskip \noindent{\em #1}}

\newcommand{\sendmessagealgorithm}{\ensuremath{{\sf SendMessage}}}
\ifdefined \sendmessage
\renewcommand{\sendmessage}{\sendmessagealgorithm}
\else
\newcommand{\sendmessage}{\sendmessagealgorithm}
\fi

\ifdefined \simulator
\renewcommand{\simulator}{\ensuremath{\mathcal{S}}}
\else

\ifdefined \negl
\renewcommand{\negl}{\ensuremath{\mathsf{negl(\securityparam)}}}
\else
\newcommand{\negl}{\ensuremath{\mathsf{negl(\securityparam)}}}
\fi

\ifdefined \distinguisher
\renewcommand{\distinguisher}{\ensuremath{\mathcal{Z}}}
\else
\newcommand{\distinguisher}{\ensuremath{\mathcal{Z}}}
\fi

\long\def\symbolfootnote[#1]#2{\begingroup%
\def\thefootnote{\fnsymbol{footnote}}\footnote[#1]{#2}\endgroup}

\let\latexcite=\cite
\def\cite{\nolinebreak\latexcite}
\let\latexref=\ref
\def\ref{\nolinebreak\latexref}

\newcommand{\ignore}[1]{}

\newcommand{\securityparam}{\ensuremath{{\lambda}}}

\newcommand{\verifyfull}{\ensuremath{\mathsf{Verify}}}
\ifdefined \verify
\renewcommand{\verify}{\verifyfull}
\else
\newcommand{\verify}{\verifyfull}
\fi

\ifdefined \adversary
\renewcommand{\adversary}{\ensuremath\mathcal{ A}}
\else
\newcommand{\adversary}{\ensuremath\mathcal{ A}}
\fi

\newcounter{romanlistcounter}
  {\setcounter{romanlistcounter}{0}%
   \begin{list}{\textit{(\roman{romanlistcounter})}}{%
        \usecounter{romanlistcounter}%
      \setlength{\itemsep}{0pc}%
      \setlength{\itemindent}{1pc}%
      \setlength{\topsep}{0pc}%
      \setlength{\mylabelwidth}{3em}}}
  {\end{list}}

\newcounter{alphalistcounter}
  {\setcounter{alphalistcounter}{0}%
   \begin{list}{\textit{(\alph{alphalistcounter})}}{%
        \usecounter{alphalistcounter}%
      \setlength{\itemsep}{0pc}%
      \setlength{\itemindent}{1pc}%
      \setlength{\topsep}{0pc}%
      \setlength{\mylabelwidth}{3em}}}
  {\end{list}}

\newcommand{\exactexpectationname}{Organization}

\begin{document}

\title{``I need a better description'':\\An Investigation Into User Expectations For Differential Privacy\thanks{A version of this paper appears in the proceedings of the 28th ACM Conference on Computer and Communications Security (CCS 2021).  Authors listed alphabetically.}}

\author[1]{Rachel Cummings}
\author[2]{Gabriel Kaptchuk}
\author[3]{Elissa M. Redmiles}
\affil[1]{Columbia University, \texttt{rac2239@columbia.edu}}
\affil[2]{Boston University, \texttt{kaptchuk@bu.edu}}
\affil[3]{Max Planck Institute for Software Systems, \texttt{eredmiles@mpi-sws.org}}


\date{}

\maketitle

\begin{abstract}
Despite recent widespread deployment of differential privacy, relatively little is known about what users think of differential privacy. In this work, we seek to explore users' privacy expectations related to differential privacy. Specifically, we investigate (1) whether users care about the protections afforded by differential privacy, and (2) whether they are therefore more willing to share their data with differentially private systems. Further, we attempt to understand (3) users' privacy expectations of the differentially private systems they may encounter in practice and (4) their willingness to share data in such systems. To answer these questions, we use a series of rigorously conducted surveys ($n=2424$).

We find that users care about the kinds of information leaks against which differential privacy protects and are more willing to share their private information when the risks of these leaks are less likely to happen.  Additionally, we find that the ways in which differential privacy is described in-the-wild haphazardly set users' privacy expectations, which can be misleading depending on the deployment. We synthesize our results into a framework for understanding a user's willingness to share information with differentially private systems, which takes into account the interaction between the user's prior privacy concerns and how differential privacy is described.
\end{abstract}

\section{Introduction}

Differential privacy (DP) is a mathematically rigorous definition of privacy that has gained popularity since its formalization in 2006 \cite{TCC:DMNS06}. DP facilitates the computation of aggregate statistics about a dataset while placing a formal bound on the amount of information that these statistics can disclose about individual data points within the dataset. Guaranteeing DP generally requires injecting carefully calibrated noise that hides individual datapoints while preserving aggregate level insights.

DP has become a leading technique used to meet the increasing consumer demand for digital privacy \cite{pewsurvey}. In the last few years, several companies have  deployed DP. For instance, Apple uses DP to gather aggregate statistics on Emoji usage, which it uses to order Emojis for users \cite{appledp, appledpshort}. Uber uses DP to prevent data analysts within the company from stalking customers \cite{uberdpconference,uberdpgithub}, and Google uses DP to crowd-source statistics from Google Chrome crash reports \cite{CCS:ErlPihKor14}.  The U.S. government has also begun to use DP. The United States Census Bureau is using DP to prevent information disclosure in the summary statistics it releases for the 2020 Decennial Census \cite{censusdp}. The use of DP in the Census means that nearly every person in the United States will have private data protected by DP. 

Following in the footsteps of these earlier adopters, more companies have already announced their intentions to integrate differentially private techniques into their systems, e.g., \cite{facebookone,facebooktwo}. As a result, DP is becoming an increasingly consumer-relevant technology. Yet, little is known about whether end users value the protections offered by DP.

While DP is mathematically elegant and computationally efficient, it can be difficult to understand. Not only is DP typically defined mathematically, the privacy protections provided by DP are not absolute and require contextualization \cite{Dwork08}. DP does not provide binary privacy (i.e., private or not private), but instead provides a statistical privacy controlled by unitless system parameters that are difficult to interpret ({\em i.e.}, the parameters $\epsilon$ and $\delta$ control the maximum amount of information that can leak about any individual entry in the dataset) \cite{TCC:DMNS06}. Additionally, DP can be deployed in different security models, and the choice of model has significant impact on the types of adversarial behavior the system can tolerate.  In the \emph{local} model, users randomly perturb their information (with the help of the collection mechanism, e.g., their device) before sending it to a central entity in charge of analysis, called the curator \cite{KLNRS11}.  In the \emph{central} model users share their sensitive information directly, and the curator is trusted to perturb results that are released \cite{DR14}. 

\medskip\noindent
\textbf{Differential Privacy from the user's perspective.} The existing DP literature focuses on techniques for achieving DP \cite{TCC:DMNS06,DR14,FOCS:McSTal07, FOCS:HarRot10,STOC:DNRRV09,KLNRS11,DFH+15}, with a small but growing body of work on legal and ethical implications of DP \cite{nissim2018privacy,cummings2018role, oberski2020differential, cohen2020towards}. Notably absent, however, is the voice of the end user, whose information may eventually be protected by DP and may benefit from its deployment~\cite{benthall2017contextual}. Do users care about the information disclosures against which DP protects? Do users understand how DP protects them, and if so, do those protections influence their comfort with sharing information? As differentially private systems proliferate, it is increasingly important to answer such questions and understand DP from the \emph{user's} perspective.
%

While a limited body of prior work has sought to understand how training users to understand DP influences willingness to share \cite{SP:XWLJ20, bullek2017towards,DCHL21}, we aim to answer broader questions regarding:  \begin{enumerate*}[(i)]
    \item whether DP meets users' existing privacy needs,
    \item what expectations a potential user might have of a differentially private system, and
    \item whether existing, in-the-wild, descriptions of DP accurately set user expectations.
\end{enumerate*}  

As a privacy-enhancing technology, DP is designed to prevent the unwanted information disclosure of user information to certain entities. However, it is not clear that these protections are meaningful to potential users. Additionally, it is not clear if DP provides the level of protection that potential users might hope.

Thus, in this work, we ask the following research questions:

\begin{enumerate}

  \item[\textbf{(RQ1)}] Do potential users care about protecting their information against disclosure to the entities against which differentially private systems can protect? 
  \item[\textbf{(RQ2)}] Are potential users more willing to share their information when they have increased confidence that such information disclosures will not occur?
  \item[\textbf{(RQ3)}] Do potential users expect differentially private systems to protect their information against disclosure? How does the way in which differential privacy is described impact their expectations?
  \item[\textbf{(RQ4)}]  Are potential users more willing to share their information when their information will be protected with differential privacy?  How does the way in which differential privacy is described impact sharing?
   
\end{enumerate}
We conduct two surveys with a total of $2,424$ respondents to answer our research questions. We use vignette-based surveys to elicit respondents intended behavior, as such surveys have been found to well-approximate real-world behavior~\cite{hainmueller2015validating}. 

To address RQ1 and RQ2, we present each respondent with one of two information-sharing scenarios (sharing information with a salary transparency initiative or sharing medical records with a research initiative) and query respondents' privacy concerns. We then set their privacy expectations for those concerns (e.g., how likely information is to be leaked to a particular entity) and query if they would be willing to share their private information. Using the results of this survey, we examine how respondents' privacy concerns align with the protections offered by DP.

To address RQ3 and RQ4, we again present each respondent with one of the two information-sharing scenarios. We additionally tell respondents that their information will be protected by DP, as described by one of six descriptions.\footnote{We also maintain a control group of participants who are not told that their information is protected.} We then query respondents' privacy expectations for these scenarios (e.g., whether their information could leak to various entities) and whether they would be willing to share their information. Using the results of this survey, we interrogate how accurately and effectively existing descriptions of DP set user expectations.

There is no ``standard'' deployment of DP, nor is there a ``normal'' way to describe its guarantees. In order to construct representative descriptions of DP to present to our participants, we systematically collected over 70 descriptions of DP written by companies, government agencies, news outlets, and academic publications. Through affinity diagramming qualitative analysis~\cite{beyer1999contextual}, we identify six main themes present in these descriptions, compose a representative description for each theme, and showed these representative definitions  to respondents.

By describing DP as a potential user would encounter it in-the-wild, we gain a better understanding of how potential users are likely to respond to DP in practice. The nuances innate in DP make it easy for a prospective user to misunderstand what they are being promised. As such, a user seeking to choose the right privacy preserving system may find it difficult to make an informed choice. Getting this wrong can have real-world consequences: DP may be insufficient to protect a user's information against the types of threats about which they are concerned.

\medskip\noindent
\textbf{Summary of Findings.} We find that users care about the kinds of information disclosure against which DP can protect (RQ1) and are more willing to share their private information when the risk of information disclosure to certain entities, specifically those for which disclosure would represent an inappropriate information flow~\cite{nissenbaum2004privacy}, is not possible (RQ2).

Further, we find that descriptions of DP raise respondent's concrete privacy expectations around information disclosure (RQ3). This effect, however, varies by how DP is described: different descriptions of DP raise expectations for different kinds of information disclosure. These expectations, in turn, raise respondent's willingness to share information. However, informing respondents that a system was differentially private did \emph{not} raise potential user's willingness to share information, no matter which description of DP was presented to the respondent (RQ4). 

Taken together, our findings suggest that while (1) respondents do care about the information disclosures against which DP can protect; (2) the likelihood of those disclosures influences respondent's willingness to share; and (3) different in-the-wild descriptions of DP influence respondents perception of the likelihood of those disclosures. However, (4) simply being shown a randomly-selected, in-the-wild description of DP does not increase willingness to share.  These results, at first glance, appear to be in tension.

On deeper analysis however, these findings suggest the presence of a misalignment between the information disclosures about which users care and the information disclosures that descriptions of DP address. The probability that a given respondent was (a) shown a description that related to the information disclosures about which they care, and (b) that description influenced enough of their perceptions is likely low. 

Synthesizing these findings, we posit a novel framework for understanding how end users reason about sharing their data under DP protections. 
Our framework --- and the findings that informed it --- offer concrete directions for reformulating DP descriptions to accurately and effectively set user's privacy expectations and increase their comfort when using differentially private systems.  Users must either be trained to carefully understand descriptions (as done in \cite{SP:XWLJ20}) or descriptions should be reformulated to directly communicate how they address the information flows that concern users (e.g., via privacy nutrition-labels~\cite{kelley2009nutrition}). If DP descriptions can be effectively reformulated, our results suggest that users may be significantly more comfortable sharing their information when given differentially private protections. 
%

    

\section{Background and Related Work}
In this section, we provide a background on DP and review prior work on communicating privacy to end users, with a specific focus on prior research studying DP-related communications.

\medskip \noindent
\textbf{Differential Privacy.} In the last decade, a growing literature on differentially private algorithms has emerged to address concerns surrounding user-level data privacy. First defined by \cite{TCC:DMNS06}, DP is a parameterized notion of database privacy that gives a mathematically rigorous worst-case bound on the maximum amount of information that can be learned about any one individual's data through the analysis of a dataset. Formally, a database $D \in \mathcal{D}^n$ is modeled as containing data from $n$ individuals, and DP constrains the change in an algorithm's output caused by changing a single person's data in the database.

\begin{definition}[Differential Privacy \protect\cite{TCC:DMNS06}]\label{def.dp}
An algorithm $\mathcal{A}: \mathcal{D}^n \rightarrow \mathcal{R}$ is \emph{$(\epsilon,\delta)$-differentially private} if for every pair of databases $D, D' \in \mathcal{D}^n$ that differ in at most one entry, and for every subset of possible outputs $\mathcal{S} \subseteq \mathcal{R}$,
\[ \Pr[\mathcal{A}(D) \in \mathcal{S}] \leq \exp(\epsilon)\Pr[\mathcal{A}(D') \in \mathcal{S}] + \delta. \]
\end{definition}

DP can be implemented either in the \emph{central model} --- where users provide their raw data to a trusted curator for private analysis --- or in the \emph{local model}, where users add noise locally to their own data before sharing it for analysis. The central model corresponds to the original DP definition of \cite{TCC:DMNS06} as presented in Definition \ref{def.dp}, where an analyst first collects a dataset from users, and then uses specialized DP tools to ensure that the technical requirements of Definition \ref{def.dp} are satisfied. The original intended use case for central DP is to enable trusted data analysts who already held sensitive datasets to publish aggregate statistics or reports on their data without violating the privacy of the individuals represented in the data. Central DP is used by e.g., the U.S. Census Bureau \cite{censusdp,censusexample} and Uber \cite{uberdpconference,uberdpgithub} since both require exact user data -- the Census Bureau through a constitutional mandate; Uber because data like rider location are necessary for their ride-sharing services.

The local model provides privacy guarantees in the presence of an untrusted curator. Users add noise to their own information (i.e., on their own device) through algorithms that satisfy Definition \ref{def.dp} for $n=1$, and share the privatized output with the curator. Thus, the curator receives only a perturbed and private version of each user's data and never has access to raw user data. Any analysis performed on the noisy data will retain the same DP guarantee due to post-processing \cite{TCC:DMNS06}, so the curator need not use any specialized analysis tools to ensure privacy. Analysts can still make aggregate inferences based on population-level statistics, but will only see noisy information about any individual. Local DP is used by, e.g., Apple \cite{appledp,appledpshort}, Google \cite{CCS:ErlPihKor14}, and Microsoft \cite{DKY17} in settings where user data is stored on-device and the company only requires aggregate information to perform its services.

The possible risks of information disclosure differ substantially between these two models. Since the central model stores user data in a centralized location, data analysts have access to exact user data, along with any other parties who obtain access through legal or illegal means. In the local model, the dataset itself is privatized, so there is no risk of information disclosure through the curator's dataset. In this work, we seek to understand user's perceptions of these possible risks of information disclosure and interrogate the accuracy of those perceptions under both the local and central models of differential privacy.
%

\medskip \noindent
\textbf{Privacy Communications.}\label{sec:sppaperbackground} A large body of work has examined how best to explain privacy to end users~\cite{mcdonald2009comparative,spiekermann2008engineering,earp2005examining,jensen2004privacy,senarath2017designing}. This has included creating privacy nutrition labels~\cite{kelley2009nutrition} that clearly delineate to users who may use their information, how their information may be used, and how likely these uses are to occur; designing privacy icons that clearly communicate when and what information is being collected~\cite{motti2016towards, egelman2015thing, cranor2006they}; and developing machine-learning systems that help users negotiate privacy boundaries~\cite{sadeh2013usable}. Particularly relevant to the work presented here, prior work has also identified best practices for privacy communications: descriptions should be relevant (e.g., include the necessary context for users to make decisions), actionable (e.g., allow the user to make choices), and understandable (e.g., usable, not overloading the user with technical information)~\cite{schaub2017designing}. As we discuss in \cref{sec:discussion}, our findings suggest that existing DP descriptions fail to satisfy these criteria.

Despite this large body of prior work on privacy communications and the increasing importance of DP, only two pieces of prior work have focused on communicating with users about DP. 

Bullek et. al. \cite{bullek2017towards} study how users understand privacy parameters in randomized response, a specific local DP technique. They describe randomized response to users using a virtual, colored spinner; the user would spin the spinner, the outcome of which would indicate if the user should answer the sensitive question truthfully or with the response indicated on the spinner. Our work focuses more broadly on how the information disclosures against which DP protects can influence users' willingness to share, and on how descriptions of DP influence expectations for those disclosures. 

Most closely related to our own work, Xiong et al. \cite{SP:XWLJ20} study how informing users that their information is protected with DP influences their willingness to share different types of information. They study this question in the context of an app that collects medically relevant information, both \emph{low sensitivity} (e.g., gender, height, weight) and \emph{high sensitivity} (e.g., substance use, income level, current medication). They found that promising users DP makes them more willing to share their information (particularly high sensitivity information, which is comparable to the information we consider in this work). However, they found that users struggled to understand descriptions of DP but were more able to understand descriptions that mentioned the implications of information sharing.

Our study builds upon this prior work to more deeply explore (RQ1) which information sharing implications are most concerning to users, and thus should be emphasized when describing DP, (RQ2) how these implications themselves influence users' willingness to share information, and (RQ3) how existing in-the-wild descriptions of DP set their expectations about these information sharing implications. Prior work on user expectations for information sharing~\cite{benthall2017contextual} notes that there is a lack of work considering how users reason about information sharing under DP. Our work fills this gap. Additionally, we seek to replicate their results through (RQ4), in which we examine how different descriptions of DP themselves influence users willingness to share information. 

In addition to primarily focusing on different research questions, our work methodologically differs from the work of Xiong et al. \cite{SP:XWLJ20} in two ways. First, we derive the descriptions of DP we use as stimuli from a systematic review of 76 in-the-wild descriptions of DP (see Section~\ref{sec:surveytwo}). In contrast, in two of their three experiments Xiong et al. \cite{SP:XWLJ20} use DP descriptions that were created to explain the definition and/or different aspects ({\em e.g.}, data perturbation) of DP to users\footnote{The most closely related definition to our work is the DP with implications description: ``To respect your personal information privacy and ensure best user experience, the data shared with the app will be processed via the differential privacy (DP) technique. That is, the app company will store your data but only use the aggregated statistics with modification so that your personal
information cannot be learned. However, your personal information may be leaked if the company’s database is compromised.''}; more similar to our approach, in their third experiment they use the descriptions from four companies that use DP, in addition to their created descriptions. Second, and relatedly, 
Xiong et al. \cite{SP:XWLJ20} test whether respondents correctly understand the implications of DP based on the description they were shown before respondents are shown additional survey questions about their willingness to share information. Respondents who did not correctly answer the understanding question(s) were presented with the DP description a second time; if they again did not understand the description, they were excluded from the study. Their results thus have important implications about how users can be educated about DP. On the other hand, this methodology does not offer insight into how users' privacy expectations or willingness to share information might be influenced by encountering descriptions of DP in-the-wild, rather than in a laboratory setting. As further discussed in Section~\ref{sec:discussion}, the results of our replication study (RQ4) significantly differ from those of Xiong et al. \cite{SP:XWLJ20} likely due to the methodological differences in our approaches.
\section{Summary of Methods}\label{sec:methods}
To answer our research questions, we ran two surveys ($n=2,424$ total), one to address RQ1 and RQ2 and the other to address RQ3 and RQ4. In order to improve the external validity of our work, we use vignette (scenario) surveys~\cite{hainmueller2015validating}. In both our surveys, we use the same two scenarios, which focus on two different potential use cases for DP: protecting people's salaries and protecting people's medical histories. To contextualize the first scenario, we ask respondents to imagine that they work in the banking industry and are approached by a friend on behalf of a salary transparency initiative. In the second scenario, we ask the respondent to imagine that during their next doctor's visit, their doctor asks them if they want to share their medical records with a medical research non-profit, in the name of improving care. The exact wording of these scenarios is shown in \cref{tab:scenariosandexpectation}.  This table also contains concrete privacy expectations about which we asked respondents in both surveys; we discuss these expectations further in the following sections. We ran both surveys using Amazon Mechanical Turk (MTurk).  MTurk has been shown to be representative of American privacy preferences for Americans aged 18-50 who have at least some college education~\cite{SP:RedKroMaz19}.

We present detailed overviews of each survey in \cref{sec:surveyone} and \cref{sec:surveytwo} respectively.  Similarly, we present our findings from each survey in \cref{sec:resultsbranch34} and \cref{sec:resultsbranch12} respectively.  Our full survey instruments can be found in \cref{apx:surveys}. We also present demographic information for our survey respondents in \cref{tab:surveydemographics}, which is also located in the appendix. Our procedures were approved by our institutions' ethics review board.

\subsection{Limitations}
As in all user studies, our study is subject to multiple possible biases. The first is sampling bias. We sample our respondents using MTurk. While prior work shows that MTurk is reasonably representative of the privacy attitudes and experiences of Americans aged 18-50 who have some college education ~\cite{SP:RedKroMaz19}, our sample does not capture the experiences of all Americans, especially those older and less educated. Our results should be interpreted in this context. Second, we may have introduced reporting biases through our question design. While we aimed to follow best practices --- using cognitive interviews to validate our questionnaire, and offering ``other'' and ``I don't know'' response options~\cite{redmiles2017summary} --- respondents may still have mis-reported or failed to report their true perceptions or preferences. Third, while we took steps to improve external validity --- sourcing experimental stimuli by rigorously collecting and coalescing in-the-wild DP descriptions and using a vignette survey --- our study may have failed to appropriately reflect real-world conditions.

\begin{table*}[!t]
\centering%
\small
\begin{tabularx}{\linewidth}{lXcc}
\toprule
\textbf{Scenario Name} & \multicolumn{3}{>{\hsize=3.1\hsize \linewidth=3.1\linewidth}X}{\textbf{Scenario Description}} \\ 
\midrule
\midrule

Salary Scenario & \multicolumn{3}{>{\hsize=10.1\hsize \linewidth=10.1\linewidth}X}{``Imagine that you work in the banking industry. You are friends with a group of other people who work in banking companies in your city. One of your friends is part of a transparency initiative that is trying to publish general statistics about pay in the banking industry. As part of this initiative, they have asked everyone in the group to share their salaries and job titles using an online web form on the initiative’s website.'' }\\
\midrule

\smallskip
Medical Scenario & \multicolumn{3}{>{\hsize=10.1\hsize \linewidth=10.1\linewidth}X}{``Imagine that during your next doctor’s visit, your primary care doctor informs you that they are part of a non-profit organization trying to push the boundaries of medical research. This non-profit is asking patients around the country to share their medical records, which will be used to help medical research on improving treatment options and patient care. Your doctor, with your permission, can facilitate the non-profit getting the information they need.''} \\
\bottomrule

\toprule
\scriptsize{\shortstack{\textbf{Expectation}\\ \textbf{Name}}} & \multicolumn{1}{>{\hsize=6.5\hsize \linewidth=6.5\linewidth}X}{\textbf{Expectation Description}} & \scriptsize{\shortstack{\textbf{Ground Truth} \\ \textbf{Local}}} & \scriptsize{\shortstack{\textbf{Ground Truth} \\ \textbf{Central}}} \\ 
\midrule
\midrule

\multirow{4}{*}{\emph{Hack}}  
&  \multicolumn{1}{>{\hsize=6.5\hsize \linewidth=6.5\linewidth}X}{``A criminal or foreign government that hacks the transparency initiative could learn my salary and job title''} \\\cline{2-2}
&  \multicolumn{1}{>{\hsize=6.5\hsize \linewidth=6.5\linewidth}X}{``A criminal or foreign government that hacks the non-profit could learn my medical history''} & False & True \\
\midrule

\smallskip 
\multirow{4}{*}{\emph{Law Enforcement}}  &  \multicolumn{1}{>{\hsize=6.5\hsize \linewidth=6.5\linewidth}X}{``A law enforcement organization could access my salary and job title with a court order requesting this data from the initiative''} \\\cline{2-2}
& \multicolumn{1}{>{\hsize=6.5\hsize \linewidth=6.5\linewidth}X}{``A law enforcement organization could access my medical history with a court order requesting this data from the non-profit''} & False & True \\
\midrule

\smallskip 
\multirow{3}{*}{\emph{\exactexpectationname}}  & \multicolumn{1}{>{\hsize=6.5\hsize \linewidth=6.5\linewidth}X}{``My friend will not be able to learn my salary and job title''} \\\cline{2-2}
& \multicolumn{1}{>{\hsize=6.5\hsize \linewidth=6.5\linewidth}X}{``The contents of my medical record will be stored only by my doctor's office, and will not be stored by the non-profit''} & True & False\\
\midrule

\smallskip 
\multirow{4}{*}{\emph{Data Analyst}}  &  \multicolumn{1}{>{\hsize=6.5\hsize \linewidth=6.5\linewidth}X}{``A data analyst working on the salary transparency initiative could learn my exact salary and job title''} \\\cline{2-2}
& \multicolumn{1}{>{\hsize=6.5\hsize \linewidth=6.5\linewidth}X}{``A data analyst working for the non-profit would be able to see my exact medical history''} & False & True \\
\midrule

\smallskip 
\multirow{4}{*}{\emph{Graphs}}  &  \multicolumn{1}{>{\hsize=6.5\hsize \linewidth=6.5\linewidth}X}{``Graphs or informational charts created using information given to the salary transparency initiative could reveal my salary and job title.''} \\\cline{2-2}
& \multicolumn{1}{>{\hsize=6.5\hsize \linewidth=6.5\linewidth}X}{``Graphs or informational charts created using information given to the non-profit could reveal my medical history.''} & False & False \\
\midrule

\smallskip 
\multirow{4}{*}{\emph{Share}}  & \multicolumn{1}{>{\hsize=6.5\hsize \linewidth=6.5\linewidth}X}{``Data that the salary transparency initiative shares with other organizations doing salary research could reveal my salary and job title''} \\\cline{2-2}
& \multicolumn{1}{>{\hsize=6.5\hsize \linewidth=6.5\linewidth}X}{``Data that the non-profit shares with other organizations doing medical research could reveal my medical history''} & False & True \\

\bottomrule
\end{tabularx}
\caption{ Scenarios (top) and information disclosure expectations (bottom) used in surveys one and two.}
\label{tab:scenariosandexpectation}
\end{table*}    
\section{Impact of Information Disclosures (RQ1 \& RQ2)}
\label{sec:surveyone}
In our first survey, we aimed to answer RQ1 and RQ2. Namely, we wanted to determine if (a) users cared about the kinds of information disclosures against which DP can protect, and (b) users would be more willing to share their sensitive information when the risk of such information disclosures decreased.

\medskip\noindent
\textbf{Information Disclosures.} 
Contextual integrity (CI) theory---a commonly used framework to explain end-user privacy reasoning---posits that users' privacy decisions depend heavily on information flows---what information is being transmitted to which entities under what privacy expectations~\cite{nissenbaum2004privacy}. Our scenarios define a set of expected information flows (e.g., salary information moving from the user, to the salary transparency initiative, under some privacy protection---as described further in the following sections). Based on our descriptions, users may have different expectations for whether unexpected information flows (e.g., information being shared with an entity they did not intend to share it with) may occur. We term these unexpected information flows ``information disclosures'' throughout the remainder of the paper. 

As survey one seeks to investigate the role of different types of information disclosures in users' sharing behaviors and survey two seeks to investigate how existing methods of describing DP influence user expectations for information disclosure, both surveys address the same potential information disclosures that could occur in either scenario.

While we would like users to tell us about the kinds of disclosure that concerns them, prior work has shown that users often do not have good mental models for privacy tools \cite{FOCI:ARUW18, SP:KBPSv19, mai2020user}. To compensate for this, we leveraged prior work~\cite{kang2015my,vitak2018privacy,morando2014privacy,carrascal2013your,dpprimer,HL14} on both DP and on user privacy concerns to create a preliminary list of information disclosures about which a user might care. Our list included the following kinds of information disclosure (names in italics): \begin{itemize*}
 \item[\emph{(Hack)}] Could a criminal organization or foreign government access the respondent's information by hacking the organization holding the information?;
 \item[\emph{(Law Enforcement)}] Could a law enforcement organization access the respondent's information with a court-issued warrant?;
 \item[\emph{(\exactexpectationname)}] Could the organization collecting the information (or their representative) access the respondent's information?;
 \item[\emph{(Data Analyst)}] Could a data analyst working within the organization access the respondent's information?;
 \item[\emph{(Graphs)}] Could graphs or informational charts created by the organization be used to learn the respondent's information?; and
 \item[\emph{(Share)}] Could the collected information be shared with another organization such that the other organization could access the respondent's information?
\end{itemize*}

While some of these questions are redundant from a technical perspective (e.g., \emph{hack} and \emph{law enforcement}), we chose these questions to be representative of real data-privacy concerns that potential users might have, since prior work finds that a key part of users' reasoning about privacy is how appropriate they consider different information flows~\cite{nissenbaum2004privacy}.

\subsection{Methodology: Survey One}\label{sec:surveyonemeth}
To ensure that we had not missed information disclosures about which users were concerned, we first conducted five cognitive interviews\footnote{Cognitive interviewing is a survey methodology technique in which participants think aloud as they answer a survey~\cite{redmiles2017summary}. Cognitive interviews are used to verify that potential respondents understand the survey questions and no answer choices are missing. We conducted interviews until no new survey protocol corrections emerged.} and offered survey respondents the opportunity to list other information disclosures about which they cared. Fewer than 2\% entered a disclosure not captured in our list. As such, we use the above list of information disclosures throughout this work. We present the descriptions of these expectations in \cref{tab:scenariosandexpectation}.

 In \cref{tab:scenariosandexpectation}, we also indicate the ground truth for each of these information disclosures in both the local and central model of DP. Both central and local DP protect against information disclosure through graphs, as this is the core privacy guarantee of DP. The central model aggregates raw user data into a centralized database that can potentially be accessed by the data analyst, employees of the organization, entities that hack the organization, law enforcement (with proper court orders), and partner organizations with whom the dataset is shared. In the local model, the aggregated dataset contains only DP versions of user data, so information disclosure would not occur even if the dataset is accessed by these entities.

We stress that we are considering a ``typical'' DP deployment and acknowledge that there are deployments for which our ground truths are not correct. For instance, we indicate that a data analyst would be able to learn a potential user's exact information in the central model. However, Uber deployed central DP specifically to protect users' information against curious data analysts. As it is impossible to account for all possible system parameters and design options, we derive our ground truth from the most simple setup.

\medskip\noindent
\textbf{Questionnaire.} Each respondent was randomly assigned to either the salary scenario or the medical scenario described above. Then, each respondent was asked to indicate which of the information disclosures, described above, they would want to better understand before sharing their information. Additionally, they were given the option of adding any additional disclosures about which they would want additional information. For each information disclosure event that the respondent indicated they would want to better understand, the respondent was presented with one of the following explicit risks, chosen at random: \begin{enumerate}[(1),leftmargin=*]
 \item there is no risk of this information disclosure,
 \item the risk of this information disclosure is the same as the chance that your bank account will be compromised (accessed by a person who you did not intend to gain access to) as part of a data breach in the next year, and 
 \item the risk of this information disclosure is higher than the chance that your bank account will be compromised (accessed by a person who you did not intend to gain access to) as part of a data breach in the next year.
\end{enumerate}
We set expectations in this way because (a) prior work on how humans interpret numbers and risk suggests that reference events of a similar type improve risk comprehension~\cite{singh1997informed,riederer2018put,gigerenzer200530,keller2009effect}, and (b) prior work shows that users have concrete estimates for the likelihood of bank account compromise, a frequently discussed security event~\cite{barrio2016improving,kaptchuk2005good,slovik1987perception}.
Each respondent was then asked if they would be willing to share their information with the initiative. Additionally, they were asked to describe why they would or would not be willing to share their data.

Finally, each survey concluded with a battery of demographic questions, including a measurement of internet skill using an existing validated measure~\cite{hargittai2012succinct}, as prior work suggests that internet skill is among the most relevant constructs to control for in privacy studies~\cite{redmiles2017digital,redmiles2018net, hargittai2013new, hargittai2019internet}. The complete survey is in \cref{apx:surveyone}.

\medskip
\noindent \textbf{Sample.} We surveyed 1,216 U.S. Amazon Mechanical Turk workers. These workers were split evenly between the two survey scenarios. To ensure high quality responses, we required that respondents have at least a 95\% approval rating~\cite{peer2014reputation}. The demographics of our sample are reported in Table~\ref{apx:surveys} in the appendix.

\medskip
\noindent \textbf{Analysis.}
To answer RQ1, we conduct a descriptive analysis, reporting the proportion of respondents who were concerned about each potential information disclosure event; when reporting differences between proportions of respondents who report concern, we use $\chi^2$ proportion tests to validate that the differences between proportions are significant. To answer RQ2, we build six logistic regression models, one for each potential information disclosure event. In each model, the dependent variable (DV) is whether the respondent is willing to share their information, the independent variable (IV) of interest is the level of risk that the respondent was told of the information disclosure occurring, and the control IVs are the scenario type and the respondents' internet skill score. We report the odds ratios (the exponentiated regression coefficients) with 95\% confidence intervals, and p-values for each IV in the model.

In \cref{sec:resultsbranch34}, we contextualize a subset of our results using open-text responses participants provided to describe their sharing decisions. These responses were analyzed through open-coding by a member of the research team with qualitative research experience. As these responses are not offered as primary research artifacts, we do not double code this data nor provide intercoder agreement statistics, per best-practice guidelines outlined in \cite{mcdonald2019reliability}.




\begin{figure}[t]
  \centering
  \includegraphics[width=0.48\textwidth]{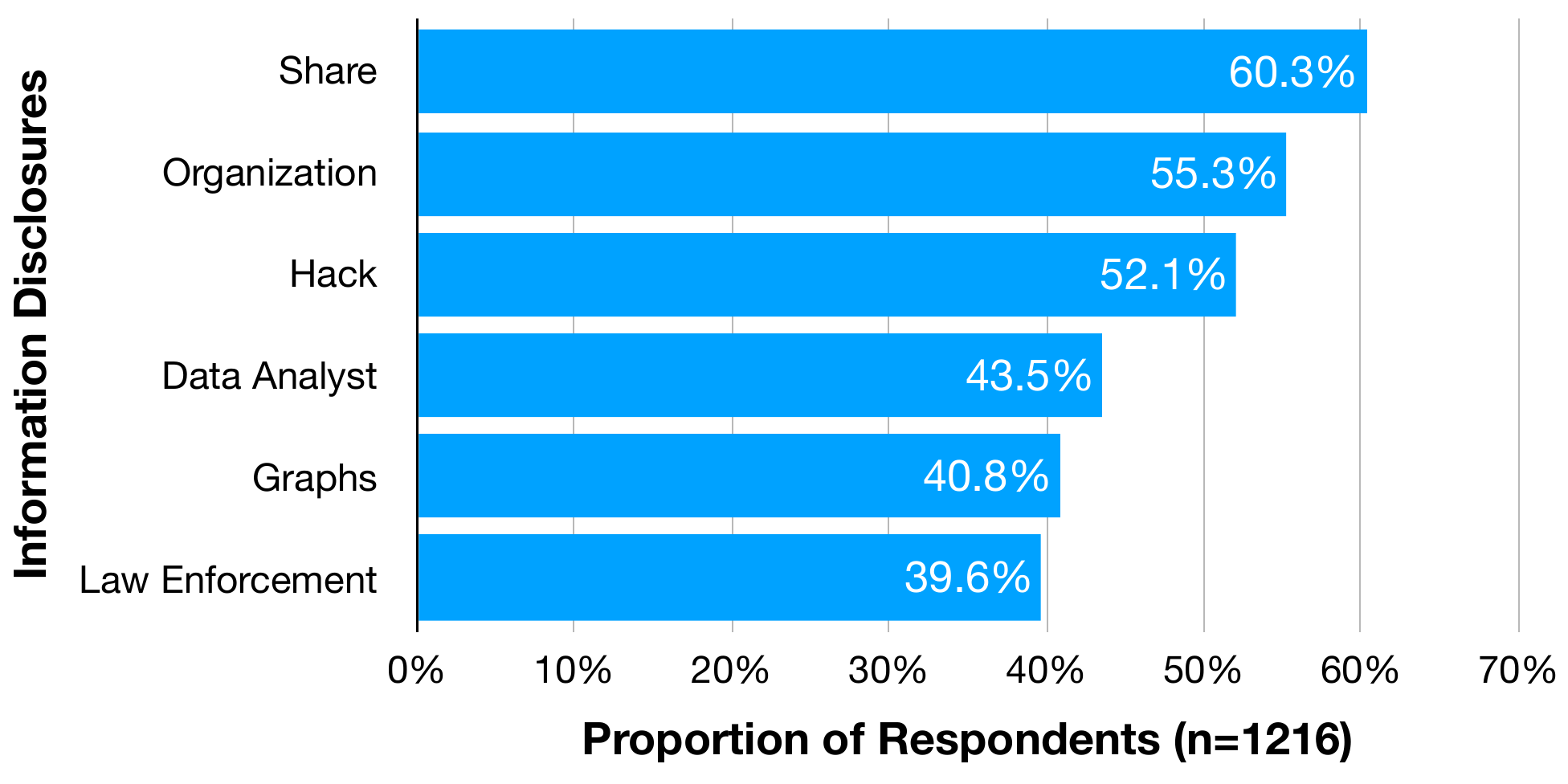}
  \caption{ Proportion of respondents who care about each potential information disclosure.}
  \label{fig:propcare}
\end{figure}

\subsection{Information Disclosure Results}\label{sec:resultsbranch34}
Here, we detail the results of our analyses of survey one.

\medskip
\noindent \textbf{RQ1: What Information Disclosures Concern Users?} 
The goal of DP is to protect user information against disclosure to various entities. Thus, we investigate whether users care about potential information disclosures to different entities against which DP can protect (see Table~\ref{tab:scenariosandexpectation} for information disclosures and  Section~\ref{sec:surveyonemeth} for the source of these disclosures).
 
We find that the most respondents --- 60.3\% --- care about information disclosures to third-parties (Share). Over half (55.3\%) care about disclosures to the person or organization running the initiative to which they contributed their information, while 52.1\% care about disclosing their information to an entity that hacks the organization to which they contributed their information. Fewer, 43.5\%, care about whether a data analyst working at the organization might be able to learn their private information or whether graphs created using their information might disclose their private information (40.8\%). Finally, 39.8\% of respondents care whether law enforcement might be able to access their information using a court order.  We visualize these findings in \cref{fig:propcare}.

\begin{table*}[t]
\centering
\small

\resizebox{\linewidth}{!}{
\begin{tabular}{lcr|cr|cr|cr|cr|cr} 
 \toprule
 \multicolumn{1}{c|}{\textbf{Variable}} & \multicolumn{2}{c|}{\textbf{\emph{Hack}}} & \multicolumn{2}{c|}{\textbf{\emph{Law Enforcement}}} & \multicolumn{2}{c|}{\textbf{\emph{\exactexpectationname}}} & \multicolumn{2}{c|}{\textbf{\emph{Data Analyst}}} & \multicolumn{2}{c|}{\textbf{\emph{Graphs}}} & \multicolumn{2}{c}{\textbf{\emph{Share}}}\\
 \midrule
  & OR/CI & p-value & OR/CI & p-value & OR/CI & p-value & OR/CI & p-value & OR/CI & p-value & OR/CI & p-value \\ 

 \midrule
 \multirow{2}{*}{Low Risk} & \textbf{1.91} & \multirow{2}{*}{\textbf{$<$ 0.01**}} & 0.87 & \multirow{2}{*}{0.55} & \textbf{1.63} & \multirow{2}{*}{\textbf{ 0.01*}} & 0.92 & \multirow{2}{*}{0.72} & 1.42 & \multirow{2}{*}{ 0.14} & 1.25 & \multirow{2}{*}{ 0.27} \\ 
& {\scriptsize [1.27, 2.88]}& & {\scriptsize [0.54, 1.39]}& & {\scriptsize [1.12, 2.38]}& & {\scriptsize [0.59, 1.44]}& & {\scriptsize [0.89, 2.27]}& & {\scriptsize [0.85, 1.84]} & \\
\multirow{2}{*}{No Risk} & \textbf{2.97} & \multirow{2}{*}{\textbf{$<$ 0.01***}} & \textbf{2.07} & \multirow{2}{*}{ \textbf{$<$ 0.01**}} & \textbf{1.61} & \multirow{2}{*}{\textbf{0.01*}} & 1.23 & \multirow{2}{*}{0.36} & 1.48 & \multirow{2}{*}{0.10} & \textbf{1.97} & \multirow{2}{*}{\textbf{$<$ 0.01***}} \\

& {\scriptsize [1.98, 4.49]}& & {\scriptsize [1.32, 3.27]}& & {\scriptsize [1.1, 2.35]}& & {\scriptsize [0.79, 1.9]}& & {\scriptsize [0.93, 2.37]}& & {\scriptsize [1.35, 2.88]} & \\
\midrule
\multirow{2}{*}{Salary Scenario} & \textbf{1.44} & \multirow{2}{*}{\textbf{0.03*}} & 1.35 & \multirow{2}{*}{0.12} & 1.20 & \multirow{2}{*}{0.23} & 1.37 & \multirow{2}{*}{0.09} & 1.12 & \multirow{2}{*}{0.54} & 1.04 & \multirow{2}{*}{0.81} \\ 
& {\scriptsize [1.04, 2]}& & {\scriptsize [0.92, 1.98]}& & {\scriptsize [0.89, 1.64]}& & {\scriptsize [0.96, 1.96]}& & {\scriptsize [0.77, 1.64]}& & {\scriptsize [0.76, 1.42]} & \\
\multirow{2}{*}{Internet Score} & 1.04 & \multirow{2}{*}{ 0.68} & 1.01 & \multirow{2}{*}{ 0.92} &0.98 & \multirow{2}{*}{ 0.79} & \textbf{1.24} & \multirow{2}{*}{\textbf{ 0.04*}} & 1.05 & \multirow{2}{*}{ 0.66} & 1.01 & \multirow{2}{*}{0.87} \\ 
& {\scriptsize [0.86, 1.25]} & & {\scriptsize [0.82, 1.24]} & & {\scriptsize [0.82, 1.16]} & & {\scriptsize [1.01, 1.52]} & & {\scriptsize [0.86, 1.28]} & & {\scriptsize [0.86, 1.2]} & \\

  \bottomrule
\end{tabular}
}
\caption{ \label{tab:explicityexpectationssharing} Effect of expectations about the probability of a disclosure on respondent willingness to share. This Table is constructed using data from Survey One. Each logistic regression model constructed only for respondents who cared about that type of disclosure, comparing against the High Risk condition. Table shows odds ratio (OR), 95\% confidence intervals for the odds ratios (shown in brackets), and p-values, where \textbf{*} indicates $p<0.05$, \textbf{**} indicates $p<0.01$, and \textbf{***} indicates $p<0.001$.}
\end{table*}

It is interesting to note that nearly 20\% more respondents ($\chi^2=34.54$, $p<0.001$) cared about their information being disclosed to a third party vs. being disclosed through graphs created using their information. Similarly, 11.8\% more respondents ($\chi^2=22.32$, $p<0.001$) cared about their information being disclosed to the organization running the initiative vs. being disclosed specifically to a data analyst at the organization. Open-answer responses offer some insight into this difference. When asked why they would (or would not) be willing to share their information, after they were told the risk of the disclosures they indicated they cared about, many respondents indicated that they believed in the cause of the initiative and wanted to contribute to their analysis/research. For example, one respondent said, ``I trust the non-profit organization to handle my information responsibly and to use it for the positive research purpose that they claim they will be using it for.'' Thus, respondents may care less about disclosures that occur through ``appropriate information flows''~\cite{nissenbaum2004privacy}, in which user's information is being used in the way they expect: e.g., to benefit salary transparency or medical research through data analysis and the generation of graphs.

\medskip
\noindent \textbf{RQ2: How Does the Probability of Information Disclosure Influence  Sharing?} 
Next, we investigate whether setting respondents' expectations about the information disclosures that concerned them influences their reported willingness to share information.  The results are shown in Table~\ref{tab:explicityexpectationssharing}.

Among respondents who care about their information being hacked by a criminal organization or foreign government, those who were told that the risk of this disclosure occurring is about the same as the risk of having their bank account compromised (``Low Risk'' in Table~\ref{tab:explicityexpectationssharing}) are nearly twice (O.R. = $1.91$, $p<0.01$) as likely to share their information as compared to respondents who were told that this risk is greater than the risk of having their bank account compromised. Those who are told there is no risk of their information being disclosed through a hack were nearly three times as likely to share (O.R. = $2.97$, $p<0.01$). Respondents who care about their information being hacked were also more willing to share salary information than medical information.

Respondents that care about information disclosure to the organization running the initiative to which they might contribute their information were $>60\%$ more willing to share if the risk of their information being disclosed to the organization was lower (Low Risk: O.R. = $1.63$, $p=0.01$; No Risk: O.R. = $1.61$, $p=0.01$).

On the other hand, respondents who care about whether law enforcement would be able to access their information with a court order and respondents who care about whether their information might be disclosed to a third-party are both more likely to share their information only if they are told there is no risk of their information being disclosed to these entities. Respondents that are told there is no such risk are about twice as likely to share their information (Law Enforcement: O.R. = $2.07$, $p<0.01$; Share: O.R. = $1.97$, $p<0.01$). Being told there is a low risk instead of a high risk of disclosure has no significant effect on their willingness to share. 

We hypothesize that respondents show a graduated response to the risk of information disclosures to the organization running the initiative to which they might contribute their information because it is appropriate for this organization to have their information. Similarly, we hypothesize that respondents show a graduated response to the risk of hacks because information disclosures resulting from hacks are unintentional on the part of the organization. On the other hand, the organization purposefully choosing to share information they contributed to the initiative with a third party or with law enforcement, even with a court order, may feel incongruent with the purpose for which they shared their information.

\begin{table*}[t!]
\centering%
\small
\begin{tabularx}{\linewidth}{lX}
\toprule
\textbf{Theme} & \textbf{Description} \\ 
\midrule
\midrule

\multirow{2}{*}{Unsubstantial} & ``Differential privacy is the gold standard in data privacy and protection and is widely recognized as the strongest guarantee of privacy available.'' \\
\midrule

\smallskip 
Techniques & ``Differential Privacy injects statistical noise into collected data in a way that protects privacy without significantly changing conclusions.'' \\
\midrule

\smallskip
\multirow{2}{*}{Enables} & ``Differential Privacy allows analysts to learn useful information from large amounts of data without compromising an individual's privacy.'' \\
\midrule

\smallskip
Trust & ``Differential privacy is a novel, mathematical technique to preserve privacy which is used by companies like Apple and Uber.''\\
\midrule

\smallskip
\multirow{2}{*}{Risk} & ``Differential privacy protects a user’s identity and the specifics of their data, meaning individuals incur almost no risk by joining the dataset.''\\
\midrule

\smallskip
\multirow{3}{*}{Technical} & ``Differential privacy ensures that the removal or addition of a single database item does not (substantially) affect the outcome of any analysis. It follows that no risk is incurred by joining the database, providing a mathematically rigorous means of coping with the fact that distributional information may be disclosive.'' \cite{Dwork08}\\

\bottomrule
\end{tabularx}
\caption{Descriptions of DP synthesized from the six main themes present in our collection of 76 in-the-wild DP descriptions.}\label{tab:descriptions}
\end{table*}

Finally, we find that the probability of disclosure to data analysts or through graphs has no effect on willingness to share, even among respondents who care about those information. We hypothesize that those respondents who are motivated by the altruistic goals specified in the scenarios may be willing to share their information regardless of the risk of disclosure occurring through these information flows, which are arguably the most appropriate information flows we examine, while those who are not compelled by the goals of the organizations described in the scenarios are similarly unwilling to share their information regardless of this risk. For example, one respondent who cared about information disclosure to both of these entities, and was told there was no risk of disclosure to a data analyst and low risk of disclosure through a graph said, ``Unfortunately, I do not see enough of a benefit for me to take the risk of sharing my personal information. I absolutely do not want such personal info being leaked out.`` On the other hand, a respondent who cared about disclosure to a data analyst and was told the risk of this disclosure was higher than the risk that their bank account would be compromised commented that they would be willing to share their information, ``because it's for good research, and I'm getting too old to worry about who sees my medical record. I anticipate I will have *many* doctors, nurses, lab techs, etc involved in my medical record before too long.''

We note that respondents who cared about information disclosures to data analysts with higher internet scores were more likely to report being willing to share their information. As technologically savvy respondents, they may have had a clearer mental model of the data analysis process and therefore understood that data analysts typically have complete access to user information. As such, they may be more forgiving toward any approach that aims to reduce this level of access, even given the relatively high risk of an information disclosure.

\begin{table*}[th!]
\centering
\small

\resizebox{\linewidth}{!}{
\begin{tabular}{lcr|cr|cr|cr|cr|cr} 
 \toprule
 \multicolumn{1}{c}{\textbf{Variable}} & \multicolumn{2}{|c}{\textbf{\emph{Hack}}} & \multicolumn{2}{|c}{\textbf{\emph{Law Enforcement}}} & \multicolumn{2}{|c}{\textbf{\emph{\exactexpectationname}}} & \multicolumn{2}{|c}{\textbf{\emph{Data Analyst}}} & \multicolumn{2}{|c}{\textbf{\emph{Graphs}}} & \multicolumn{2}{|c}{\textbf{\emph{Share}}}\\
 \midrule
  & OR/CI & p-value & OR/CI & p-value & OR/CI & p-value & OR/CI & p-value & OR/CI & p-value & OR/CI & p-value \\ 
 \midrule
 Description: & \textbf{1.94} & \multirow{2}{*}{\textbf{0.01*}} & 1.10 & \multirow{2}{*}{0.72} & 1.13 & \multirow{2}{*}{0.59} & 1.71 & \multirow{2}{*}{0.10} & \textbf{1.64} & \multirow{2}{*}{\textbf{0.05*}} & 1.68 & \multirow{2}{*}{0.06} \\ 
~~ \emph{Unsubstantial} & {\scriptsize [1.16, 3.29]} & & {\scriptsize [0.65, 1.86]} & & {\scriptsize [0.73, 1.75]} & & {\scriptsize [0.92, 3.27]} & & {\scriptsize [1.01, 2.67]} & & {\scriptsize [0.99, 2.88]} & \\
Description: & \textbf{1.96} & \multirow{2}{*}{\textbf{0.01*}} & 1.21 & \multirow{2}{*}{0.47} & 1.43 & \multirow{2}{*}{0.10} & \textbf{2.40} & \multirow{2}{*}{\textbf{$<$ 0.01**}} & \textbf{2.15} & \multirow{2}{*}{\textbf{$<$ 0.01**}} & \textbf{2.22} & \multirow{2}{*}{\textbf{$<$ 0.01**}} \\ 
~~ \emph{Techniques} & {\scriptsize [1.17, 3.33]} & & {\scriptsize [0.72, 2.03]} & & {\scriptsize [0.93, 2.22]} & & {\scriptsize [1.33, 4.5]} & & {\scriptsize [1.34, 3.5]} & & {\scriptsize [1.33, 3.77]} & \\
Description: & 1.60 & \multirow{2}{*}{0.08} & 1.05 & \multirow{2}{*}{0.84} & 1.40 & \multirow{2}{*}{0.13} & \textbf{2.06} & \multirow{2}{*}{\textbf{0.02*}} & \textbf{1.76} & \multirow{2}{*}{\textbf{0.02*}} & 1.69 & \multirow{2}{*}{0.05} \\ 
~~ \emph{Enables} & {\scriptsize [0.95, 2.73]} & & {\scriptsize [0.63, 1.77]} & & {\scriptsize [0.91, 2.16]} & & {\scriptsize [1.13, 3.88]} & & {\scriptsize [1.09, 2.87]} & & {\scriptsize [1, 2.9]} & \\
Description: & \textbf{1.86} & \multirow{2}{*}{\textbf{0.02*}} & 1.04 & \multirow{2}{*}{0.89} & 1.43 & \multirow{2}{*}{0.11} & \textbf{1.99} & \multirow{2}{*}{\textbf{0.03*}} & 1.38 & \multirow{2}{*}{0.20} & 1.19 & \multirow{2}{*}{0.55} \\ 
~~ \emph{Trust} & {\scriptsize [1.11, 3.17]} & & {\scriptsize [0.61, 1.76]} & & {\scriptsize [0.92, 2.22]} & & {\scriptsize [1.08, 3.78]} & & {\scriptsize [0.84, 2.28]} & & {\scriptsize [0.68, 2.09]} & \\
Description: & \textbf{2.58} & \multirow{2}{*}{\textbf{$<$ 0.01***}} & \textbf{1.86} & \multirow{2}{*}{\textbf{0.01*}} & 1.43 & \multirow{2}{*}{0.10} & \textbf{2.46} & \multirow{2}{*}{\textbf{$<$ 0.01**}} & \textbf{2.40} & \multirow{2}{*}{\textbf{$<$ 0.01***}} & \textbf{2.27} & \multirow{2}{*}{\textbf{$<$ 0.01**}} \\ 
~~\emph{Risk} & {\scriptsize [1.57, 4.33]} & & {\scriptsize [1.15, 3.05]} & & {\scriptsize [0.93, 2.20]} & & {\scriptsize [1.37, 4.59]} & & {\scriptsize [1.50, 3.88]} & & {\scriptsize [1.37, 3.84]} & \\
Description: & 1.56 & \multirow{2}{*}{0.10} & 1.02 & \multirow{2}{*}{0.95} & 1.38 & \multirow{2}{*}{0.15} & \textbf{2.30} & \multirow{2}{*}{\textbf{0.01**}} & \textbf{1.70} & \multirow{2}{*}{\textbf{0.03*}} & \textbf{1.90} & \multirow{2}{*}{\textbf{0.02*}} \\ 
~~\emph{Technical} & {\scriptsize [0.92, 2.69]} & & {\scriptsize [0.60, 1.73]} & & {\scriptsize [0.89, 2.14]} & & {\scriptsize [1.26, 4.33]} & & {\scriptsize [1.04, 2.79]} & & {\scriptsize [1.12, 3.25]} & \\
\midrule
\multirow{2}{*}{Salary Scenario} & \textbf{1.32} & \multirow{2}{*}{\textbf{0.04*}} & 0.80 & \multirow{2}{*}{0.11} & \textbf{1.29} & \multirow{2}{*}{\textbf{0.03*}} & 0.75 & \multirow{2}{*}{0.06} & 1.23 & \multirow{2}{*}{0.10} & 1.14 & \multirow{2}{*}{0.31} \\ 
& {\scriptsize [1.02, 1.71]} & & {\scriptsize [0.61, 1.05]} & & {\scriptsize [1.03, 1.63]} & & {\scriptsize [0.56, 1.01]} & & {\scriptsize [0.96, 1.57]} & & {\scriptsize [0.88, 1.49]} & \\
\multirow{2}{*}{Internet Score} & \textbf{1.17} & \multirow{2}{*}{\textbf{0.04*}} & \textbf{1.25} & \multirow{2}{*}{\textbf{0.01**}} & 1.02 & \multirow{2}{*}{0.78} & 1.05 & \multirow{2}{*}{0.54} & 1.07 & \multirow{2}{*}{0.36} & 1.01 & \multirow{2}{*}{0.85} \\ 
& {\scriptsize [1.01, 1.36]} & & {\scriptsize [1.06, 1.46]} & & {\scriptsize [0.89, 1.17]} & & {\scriptsize [0.89, 1.25]} & & {\scriptsize [0.93, 1.23]} & & {\scriptsize [0.87, 1.18]} & \\

  \bottomrule

\end{tabular}
}
\caption{\label{tab:betterbydefn} Effect of DP descriptions on respondent's perception of the likelihood that their information will be disclosed through a particular information flow.  All models are logistic regressions constructed using data from Survey Two. See Table~\ref{tab:explicityexpectationssharing} for detailed legend.
}
\end{table*}


\section{Expectations \& Willingness to Share Under DP (RQ3 \& RQ4)}
\label{sec:surveytwo}
Next -- via a second survey -- we explore how DP influences privacy expectations (RQ3) and intent to share information (RQ4).

\medskip\noindent
\textbf{Descriptions of Differential Privacy.}
In order to answer these research questions, we needed to describe DP to respondents in our surveys. However, there is no standard description of DP we can use. Because we want to ask our research questions in a realistic context, we seek to describe DP to our respondents in the same way they might encounter DP in-the-wild.

To determine how DP is described in-the-wild, we conducted a systematic search for publicly available descriptions of DP using keywords such as ``differential privacy,'' ``formal privacy,'' ``privacy guarantee,'' and ``census privacy.''  We used both Google search and searched within the past five years (2014-2019) of content in large media venues. We continued searching until new search results stopped appearing. We put special focus on collecting descriptions used by industry and in the media coverage, as these descriptions are the ones that an uninformed consumer would be most likely to encounter. We performed this search and data collection in December 2019.\footnote{Since we conducted this survey, more companies and organizations have started adopting and publicly writing about DP. As such, our dataset is no longer comprehensive. Because this data collection informed the design of our survey, we choose not to incorporate the newer descriptions into our dataset.} In total, we collected 76 descriptions of DP: 36 from industry, 30 from media outlets, and 10 from the academic literature. 

The industry descriptions primarily came from companies that use DP, including Google, Apple, Microsoft, and Uber, as well as smaller start-ups and an investment firm. We also gathered multiple descriptions from the U.S. Census Bureau regarding the use of DP in the 2020 Census. The media descriptions were from large, mainstream media outlets, such as The New York Times, Fox News, The Washington Post, The Guardian, and Tech-Crunch. The academic descriptions were collected from some of the most-cited papers and books on DP, e.g., \cite{Dwork08}. As DP is an active area of research, these ten academic descriptions are clearly not comprehensive, but serve as a representative example of academic descriptions. 

 The research team employed affinity diagramming~\cite{beyer1999contextual} to extract the main themes of these widely varying descriptions of differential privacy.  In affinity diagramming, the research team collaboratively sorts pieces of content --- in our case the descriptions of DP --- into themes based on affinity, with each researcher iterating over the affinity diagram at least twice until consensus was reached on appropriate categorization. This analysis resulted in the identification of six main themes (names in bold): \begin{enumerate*}
  \item[\textbf{(Unsubstantial)}] claims that DP is the best notion of privacy;
  \item[\textbf{(Techniques)}] explanations that briefly summarize the methods used to create differentially private summary statistics, usually focusing on statistical noise;
  \item[\textbf{(Enables)}] statements that attempt to capture the types of applications that DP makes possible;
  \item[\textbf{(Trust)}] descriptions that focus on the well known organizations and companies that have recently started using DP;
  \item[\textbf{(Risk)}] statements that highlight the data-privacy risks that an individual incurs when allowing their information to be part of a differentially private system; and 
  \item[\textbf{(Technical)}] highly technical explanations using dense, mathematical language.
\end{enumerate*}

Many of the descriptions we gathered touch on more than one of these main themes. For instance, documents prepared by the U.S. Census Bureau state, ``Differential privacy allows us to inject a precisely calibrated amount of noise into the data to control the privacy risk of any calculation or statistic'' \cite{censusexample}. This description touches on the \emph{techniques} theme and the \emph{risk} theme, while also using \emph{technical} language like ``calculation or statistic'' that may be unnatural to non-experts. The New York Times provides a description that is another combination of the main themes, writing, ``[o]ne example, differential privacy, is already used by Apple, Google and even the U.S. Census Bureau to limit the amount of personal information that is shared with an organization while still allowing it to make useful inferences from the data'' \cite{nytexample} This description contains elements of both the \emph{trust} theme and the \emph{enables} theme.

We note that most descriptions we gathered did nothing to distinguish between the central model and the local model. Indeed, we found that determining if an industry system was in the local or central model generally required looking at technical documentation. The descriptions provided by media coverage also generally did not include any indication as to the model of system being described.

\subsection{Methodology: Survey Two}
After collecting and analyzing the descriptions of DP used in practice, we distilled six descriptions of DP that were representative of the descriptions in each of these themes. We present these descriptions in \cref{tab:descriptions}. Each of these descriptions is a synthetic creation meant to be representative of the real descriptions we found, with the exception of the \emph{technical} description, which was taken from \cite{Dwork08}. We chose to create new descriptions (rather than selecting a representative example) in order to free the description from the surrounding context, make them consistent in their structure, and have each description focus on only one theme. Just like the descriptions we observed in-the-wild, our descriptions included no indication as to whether the system they described was in the local or central model.
\smallskip 

\noindent \textbf{Questionnaire.} First, as in survey one, each respondent was randomly presented one of the two scenarios. Each respondent was then randomized to either a control condition, where there was no mention of privacy protection, or shown one of the DP descriptions. In the later conditions, the scenario was followed by the description, "To reduce the intrusion into personal privacy, the [organization] will use a technique called differential privacy. [differential privacy description]," where the description presented was sampled with equal weight from \cref{tab:descriptions}. 
Respondents were then asked the following questions: First, they were asked if they would be willing to share their data. Next, each respondent was asked to share their concrete privacy expectations by reporting whether they believed the expectations described in \cref{tab:scenariosandexpectation} (e.g., ``A criminal or foreign government that hacks the transparency initiative could learn my salary and job title.'') were true or false.   All questions included an "I don't know" option. Finally, we included the same demographics questions as above.  The complete survey is in \cref{apx:surveytwo}.

\medskip\noindent
\textbf{Sample.} We surveyed 1,208 Amazon Mechanical Turk workers following the same screening requirements as in survey one, as described in Section~\ref{sec:surveyone}. Workers were split evenly between the two survey scenarios and equally between the seven description conditions (six descriptions of DP and the baseline). 

\medskip\noindent
\textbf{Analysis.}
To answer RQ3, we construct six logistic regression models~\footnote{We note that these models are not corrected for multiple testing in line with \cite{benjamini1995controlling}, which suggests such correction only for models with a large number of DVs.}, one for each information disclosure event. The DV is whether the respondent reported that they thought the given information disclosure would occur, the IV of interest is the description they were shown (a categorical variable with the control -- no description shown -- as the baseline), the control IVs are, as in Section~\ref{sec:surveyone}, whether the scenario was the salary scenario and the respondent's internet skill. To answer RQ4, we construct a single logistic regression model. The DV is whether the respondent reported that they would be willing to share their information, the IV of interest is the description they were shown (coded as above), and the control IVs are the same as above.

\begin{figure*}[h!]
  \centering
  \subfloat[][\small \label{fig:dist:correct:local}Distribution of respondent correctness about information disclosures under local DP. The x-axis shows percentage of disclosures for which their expectations were correct under local DP.]{\includegraphics[width=.47\textwidth]{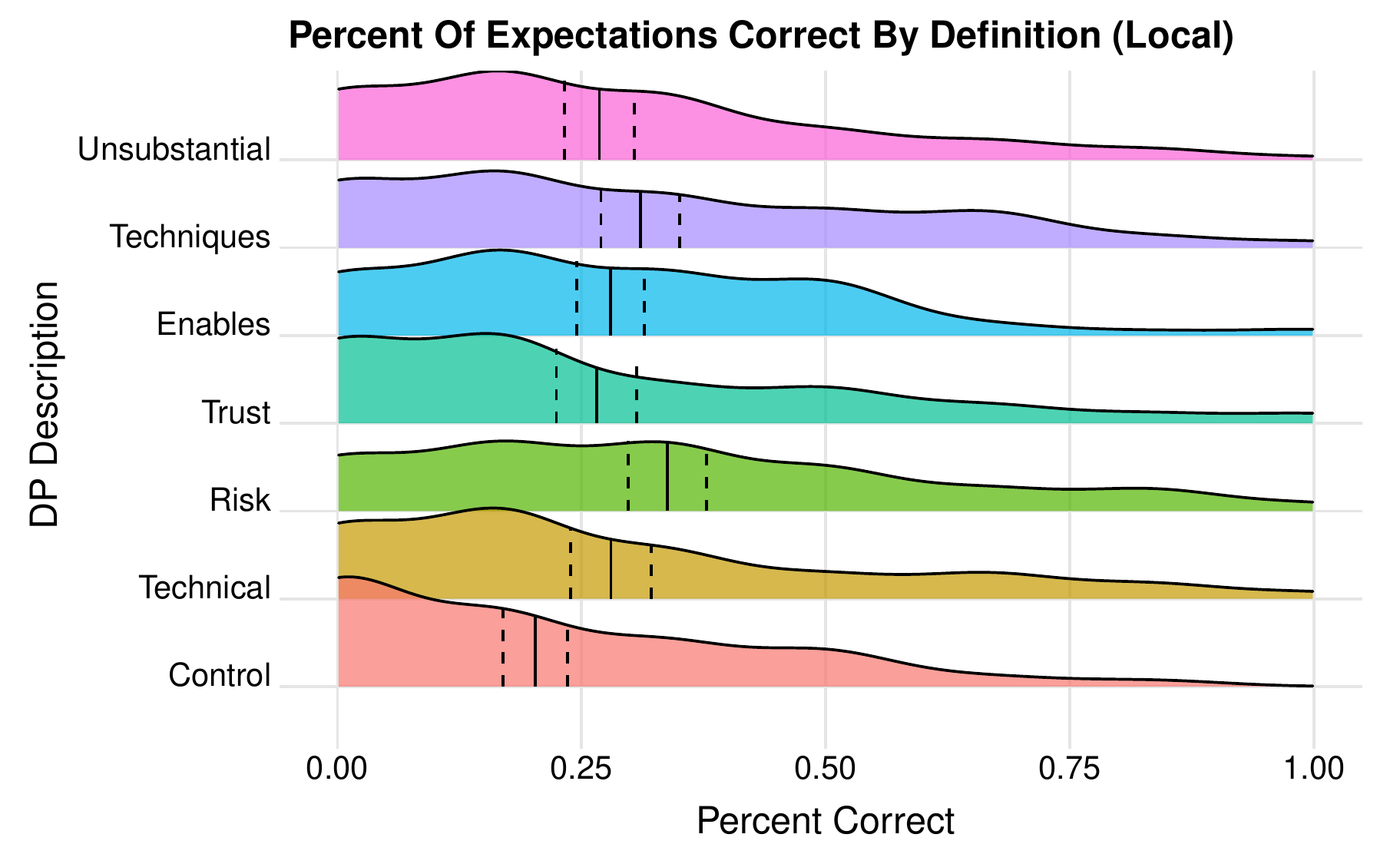}
  } \hfill
  \subfloat[][\small \label{fig:dist:correct:central}Distribution of respondent correctness about information disclosures under central DP. The x-axis shows percentage of disclosures for which their expectations were correct under central DP.]{ \includegraphics[width=.47\textwidth]{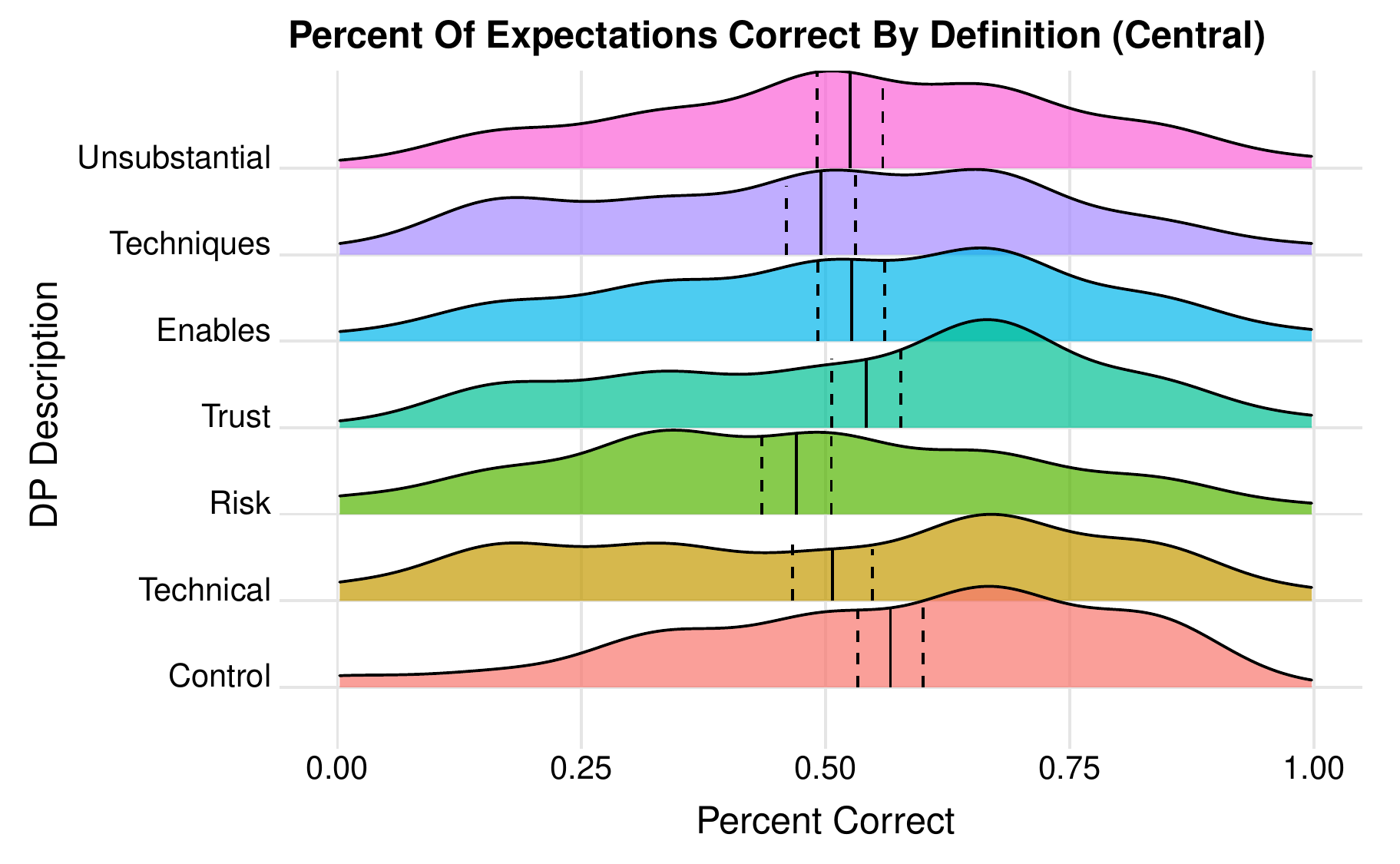}}
  \caption{ Influence of in-the-wild DP descriptions on respondent expectations for information disclosures by DP model. Dashed lines are 95\% confidence intervals of the distribution's mean. ``I don't know'' not included as correct.}
  \label{fig:distributions}
\end{figure*}

\subsection{Results: Responses to Descriptions of DP} \label{sec:resultsbranch12}
Here, we detail the results of our analyses of survey two.

\medskip\noindent
\textbf{RQ3: How Do Differential Privacy Descriptions Affect Privacy Expectations?} We divide our results for RQ3 into two parts. First, we detail our findings regarding the way descriptions of DP \emph{increase} respondents' privacy expectations (see \cref{tab:betterbydefn}). Second, we investigate if the descriptions \emph{correctly} set respondents' privacy expectations, with respect to the ground truth privacy properties of typical local and central DP deployments (see \cref{fig:distributions}).

\medskip\noindent
\textit{Descriptions Increasing Expectations.} Overall, we find that each description of DP that we tested increased respondents' privacy expectations for at least one of the disclosure risks. However, different descriptions increased different privacy expectations. 


First, we found that none of the descriptions significantly changed respondents' expectations when it came to disclosing their information to the organization soliciting their information it's representative. Respondents had higher privacy expectations in the salary scenario than in the medical scenario, indicating that the slightly different wording of these expectations may have had an effect on respondent expectations. We verified that this wording did not interfere with our main finding (that no descriptions increased user's expectations of the \emph{\exactexpectationname} disclosure) by re-building our models on each dataset separately; we found the same results.

Four of the descriptions do, however, influence respondents' perceptions regarding whether their information could be disclosed through a hack. Respondents who were shown the \emph{Unsubstantial}, \emph{Techniques}, and \emph{Trust} explanations were nearly two times more likely (\emph{Unsubstantial:} O.R. = 1.94, $p = 0.01$, \emph{Techniques:} O.R. = 1.96, $p = 0.01$, \emph{Trust:} O.R. = 1.86, $p$ = 0.02) to think their information would not be disclosed through a hack. Those who were shown the \emph{Risk} description were even more likely (O.R. = 2.58, $p < 0.01$) to think their information could not be disclosed through a hack. Users may see preventing hacks as one of the key roles of security and privacy technologies, as these results indicate that they expect such protection from the gold standard techniques (\emph{Unsubstantial}) and those used by major companies (\emph{Trust}). The \emph{Risk} description directly addresses this potential information disclosure, so it is unsurprising that it raised privacy expectations regarding hacks. Finally, respondents may have gathered that injecting statistical noise (\emph{Techniques}) would protect their information against hacks, as it does in practice.

Only the \emph{Risk} description significantly influenced respondents' perceptions of whether their information would be disclosed to law enforcement as the result of a court order: those who saw the \emph{Risk} description were nearly two times more likely (O.R. 1.86, $p$ = 0.01) to think their information would not be disclosed to law enforcement. Interestingly, this indicates that users may see information sharing with law enforcement as a risk, rather than an information flow that is appropriate and necessary to protect society. 

All of the descriptions aside from the \emph{Unsubstantial} description increase the likelihood that respondents think their information would be secure against disclosures to a data analyst, while all but the \emph{Trust} description increase the likelihood that respondents think their information would not be disclosed through graphs or charts made using their information. It may be that the \emph{Unsubstantial} description did not raise respondents' expectations for disclosure to data analysts because users are unfamiliar with the notion that data analysts could accomplish their job without seeing user information --- users may expect even ``gold standard'' techniques to disclose information in this way. Similarly, users may be unfamiliar with the idea that tech companies create graphs and charts, as most of these releases are not customer facing. Therefore, it would not be assumed that such techniques could protect user information.

The \emph{Techniques}, \emph{Risk}, and \emph{Technical} descriptions all increase the likelihood that respondents think their information could not be shared with another organization (\emph{Techniques:} O.R. 2.22, $p < 0.01$, \emph{Risk:} O.R. 2.27, $p < 0.01$, \emph{Technical:} O.R. 1.90, $p = 0.02$). As above, users may have gathered that the injection of statistical noise described in the \emph{Techniques} description would prevent this information disclosure. Additionally, both the \emph{Risks} and \emph{Technical} descriptions speak to the risk of joining the dataset. As indicated in our results for RQ1, a large number of respondents care about their information being disclosed to a third party. This may be a primary ``risk'' in their mind, which the descriptions suggest that they would be secure against.

\noindent \textit{Descriptions Setting Expectations Correctly.} Not all DP techniques reduce the likelihood of all potential information disclosures. It is critical that descriptions of DP are used to set users' expectations correctly, not only raise expectations. This is especially important in DP; a potential user encountering a description of a DP may set their expectations as though the system offers local DP, only to discover later that their information was more vulnerable because the deployment used central DP 
(see \cref{tab:scenariosandexpectation} for the ground truth we consider under both local and central DP). 

Revisiting our findings above, we note that local DP provides protection against all of the information disclosure risks about which we asked. This is because under local DP the curator never has access to the unperturbed data (and therefore cannot accidentally or intentionally disclose information). As such, increased expectations mean more correct expectations under local DP. As we saw above, each of the descriptions does increase some --- but not all --- expectations. This means that the descriptions are not only raising user expectations for differentially private systems, but setting those expectations more accurately for local privacy.

In central DP, on the other hand, the curator has access to users' raw information. In this model, the curator is responsible for injecting statistical noise into any aggregations that are released for public consumption. Because the curator has access to raw information, a typical deployment would be able to disclose sensitive information in all of the listed ways, with the exception of \emph{Graphs}. Thus, the descriptions that raise privacy expectations related to \emph{Hack}, \emph{Law Enforcement}, \emph{Data Analyst}, and \emph{Share} disclosures are actually misleading users 
in the central model.

Finally, we also consider the aggregate effect of the descriptions we study on the accuracy of respondents' privacy expectations (see \cref{fig:distributions}). We find that respondents' expectations of DP are more in line with the central model than the local model (ie. they have lower privacy expectations). Specifically, respondents had correctly set expectations for less than half of the information disclosure risks under local DP, while roughly half of their expectations were set correctly for the central model. More importantly, we note that users' privacy expectations are poorly aligned with \emph{both} local and central DP. This indicates that users have no coherent mental model of the data collection process, as many of the privacy expectations about which we ask are equivalent from a technical perspective.
\begin{table}[t!]
\centering

\small
\begin{tabular}{lrll}
 \toprule
 Variable & Odds Ratio & CI & p-value \\ 
 \midrule
 Description: \emph{Unsubstantial}	& 1.22 & [0.79, 1.88] & 0.37 \\
 Description: \emph{Techniques} 	& 0.96 & [0.62, 1.47] & 0.83 \\
 Description: \emph{Enables} 	 & 1.48 & [0.96, 2.29] & 0.08 \\
 Description: \emph{Trust} 		 	& 1.08 & [0.7, 1.67] & 0.72 \\
 Description: \emph{Risk} 			& 1.37 & [0.89, 2.12] & 0.15 \\
 Description: \emph{Technical} 		& 0.94 & [0.61, 1.45] & 0.77 \\
 \midrule
Salary Scenario & \textbf{1.67} & \textbf{[1.32, 2.1]} & $<$ \textbf{0.01***} \\ 
Internet Score & 1.09 & [0.95, 1.25] & 0.2 \\
  \bottomrule
\end{tabular}
\caption{\label{tab:sharewithdp} Effect of DP descriptions on respondents' likelihood of being willing to share information.  See Table~\ref{tab:explicityexpectationssharing} for detailed legend.
}
\end{table}

\medskip\noindent
\textbf{RQ4: How Do In-The-Wild Descriptions of Differential Privacy Affect Sharing?} 
When analyzing the results of our second survey, we find that respondents who were told that their information would be protected by DP techniques, as described by one of six different descriptions of those techniques, were no more likely to report that they would share their information in either scenario (Table~\ref{tab:sharewithdp}). We also note that the descriptions did not decrease the likelihood that respondents would be willing to share their information. Respondents were, however, more likely (O.R. = 1.67, $p < 0.01$) to share their information in the salary scenario than in the medical scenario, in line with prior work suggesting that medical information is particularly sensitive~\cite{ion2011home}. 

We note that this finding contradicts the findings of \cite{SP:XWLJ20}, who found that DP increased respondents' willingness to share high sensitivity information. We note that (a) the ways in which we describe DP and (b) the context in which we elicit responses are different. As discussed in \cref{sec:sppaperbackground}, the descriptions used in this prior work were significantly longer and not necessarily representative of in-the-wild descriptions from which we derived our descriptions, and the methodology of prior work also involved mechanisms to ensure respondents correctly understood the privacy guarantees detailed in the descriptions.

\begin{figure*}[t]
	\centering
	\resizebox{\textwidth}{!}{
		\begin{tikzpicture}
		
		\node[anchor=west] at (-13,6) {Legend:};
 \node[circle,draw,fill=red!40] at (-12.5, 5.5) {};
 \node[circle,draw,fill=orange!40] at (-12.5, 5) {};
 \node[circle,draw,fill=yellow!40] at (-12.5, 4.5) {};
 \node[circle,draw,fill=green!40] at (-12.5, 4) {};
 \node[circle,draw,fill=blue!40] at (-12.5, 3.5) {};
 \node[circle,draw,fill=pink!40] at (-12.5, 3) {};
 
 \node[anchor=west] at (-12.25, 5.5) {Hack};
 \node[anchor=west] at (-12.25, 5) {Law Enforcement};
 \node[anchor=west] at (-12.25, 4.5) {\exactexpectationname};
 \node[anchor=west] at (-12.25, 4) {Data Analyst};
 \node[anchor=west] at (-12.25, 3.5) {Graphs};
 \node[anchor=west] at (-12.25, 3) {Share};


 \node[alice,minimum size=1cm] at (-10,.5) {};
 \node[] at (-6,.5) {Description: \emph{Enables}};
 
 \node at (-10,-.5) {User Concerns:};
 \node[circle,draw,fill=pink!40] at (-9.5, -1) {};
 \node[circle,draw,fill=orange!40] at (-10, -1) {};
 \node[circle,draw,fill=red!40] at (-10.5, -1) {};

 \node at (-6,-.5) {Description Effects:};
 \node[circle,draw,fill=green!40] at (-6.25, -1) {};
 \node[circle,draw,fill=blue!40] at (-5.75, -1) {};

 \draw [->] (-10,1.25) -- (-9,2.125);
 
 \draw [->] (-6,1.25) -- (-7,2.125);
	 
 \node[circle,draw,minimum size=70] at (-7.5, 3.5) {};
 \node[circle,draw,minimum size=70] at (-8.5, 3.5) {};
 \node at (-8,1.75) {No Overlap};

 \node[circle,draw,fill=red!40] at (-9, 4) {};
 \node[circle,draw,fill=pink!40] at (-9, 3) {};
 \node[circle,draw,fill=orange!40] at (-9.25, 3.5) {};
 \node[circle,draw,fill=blue!40] at (-7, 3) {}; 
 \node[circle,draw,fill=green!40] at (-7, 4) {};
 
 \draw[->] (-8,5) -- (-8,6) {};
 \node at (-8,6.25) {No Increased Sharing};


 \node[bob,minimum size=1cm] at (-2,.5) {};
 \node at (2,.5) {Description: \emph{Risk}};
 
 \node at (-2,-.5) {User Concerns:};
 \node[circle,draw,fill=blue!40] at (-1.75, -1) {};
 \node[circle,draw,fill=yellow!40] at (-2.25, -1) {};

 \node at (2,-.5) {Description Effects:};
 \node[circle,draw,fill=red!40] at (1, -1) {};
 \node[circle,draw,fill=orange!40] at (1.5, -1) {};
 \node[circle,draw,fill=green!40] at (2, -1) {};
 \node[circle,draw,fill=blue!40] at (2.5, -1) {};
 \node[circle,draw,fill=pink!40] at (3, -1) {};
 
 \draw [->] (-2,1.25) -- (-1,2.125);
 
 \draw [->] (2,1.25) -- (1,2.125);
 
 \node[circle,draw,minimum size=70] at (-.5, 3.5) {};
 \node[circle,draw,minimum size=70] at (.5, 3.5) {};
 \node at (0,1.75) {Partial Overlap};
 
 \node[circle,draw,fill=blue!40] at (0, 3.5) {};
 \node[circle,draw,fill=yellow!40] at (-1.25, 3.5) {};
 \node[circle,draw,fill=red!40] at (1, 4.25) {};
 \node[circle,draw,fill=orange!40] at (1.25, 3.75) {};
 \node[circle,draw,fill=green!40] at (1.25, 3.25) {};
 \node[circle,draw,fill=pink!40] at (1, 2.75) {};

 \draw[->] (0,5) -- (0,6) {};
 \node at (0,6.25) {Some Increased Sharing};
 
 
 \node[charlie,minimum size=1cm] at (6,.5) {};
 \node at (10,.5) {Description: \emph{Techniques}};
 
 \node at (6,-.5) {User Concerns:};
 \node[circle,draw,fill=green!40] at (5.5, -1) {};
 \node[circle,draw,fill=blue!40] at (6, -1) {};
 \node[circle,draw,fill=pink!40] at (6.5, -1) {};

 \node at (10,-.5) {Description Effects:};
 \node[circle,draw,fill=green!40] at (9.5, -1) {};
 \node[circle,draw,fill=blue!40] at (10, -1) {};
 \node[circle,draw,fill=pink!40] at (10.5, -1) {};

 \draw [->] (6,1.25) -- (7,2.125);
 
 \draw [->] (10,1.25) -- (9,2.125);
 
 \node[circle,draw,minimum size=70] at (7.5, 3.5) {};
 \node[circle,draw,minimum size=70] at (8.5, 3.5) {};
 \node at (8,1.75) {Full Overlap};
 
 \draw[->] (8,5) -- (8,6) {};
 \node at (8,6.25) {Significantly Increased Sharing};

 \node[circle,draw,fill=green!40] at (8, 4) {};
 \node[circle,draw,fill=blue!40] at (8, 3.5) {};
 \node[circle,draw,fill=pink!40] at (8, 3) {};

		\end{tikzpicture}
	}
	\caption{{ Visualization of our framework for reasoning about the impact of DP descriptions on users' willingness to share information. Colored dots under users represent information flows about which that user cares. Colored dots under descriptions represent the information flows for which the description raises expectations. We imagine a potential user with some prior set of information disclosures about which they are concerned. When asked if they are willing to share their information with a differentially private system, the user is given a brief description of DP. Our results suggest that a user's willingness to share information is not simply a function of how this description raises their expectations, but also a function of their prior concerns. Specifically, the description may raise the user's expectations for information disclosures about which the user was not concerned. Thus, we speculate that the degree to which the user's expectations overlap with the effects of a description will be an important determining factor in a user's willingness to share.}}
	\label{fig:framework}
\end{figure*}
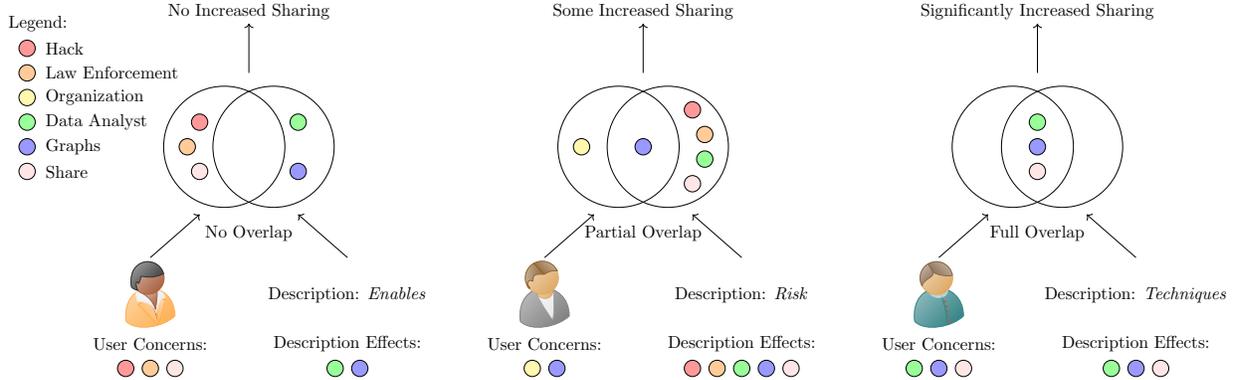

\section{Discussion}\label{sec:discussion}
\noindent
\textbf{Summary of Findings.} Our surveys indicate that (RQ1) users care about the kind of information disclosures against which DP can protect, and (RQ2) users' willingness to share information is significantly related to the degree of risk of most information disclosures occurring. However, the risk of disclosures occurring through the two information flows that might seem most appropriate~\cite{nissenbaum2004privacy} to users given our scenarios related to a salary transparency initiative and a medical research initiative --- disclosures through graphs or to a data analyst --- did not significantly relate to users' willingness to share. This is noteworthy, as ensuring privacy in graphs and informational charts is a common motivating example of DP, is the only information disclosure protected against by both local and central DP, and at least one current deployment of DP is focused on protecting user information from data analysts \cite{uberdpconference}.

We also find that in-the-wild descriptions of DP have a substantial impact on user privacy expectations (RQ3), but not user willingness to share (RQ4). Descriptions of DP that focus on different themes raise privacy expectations for information flows. However, this can be a double-edged sword, as raising expectations can also mislead users about the privacy properties of a system.

\medskip\noindent
\textbf{Novel framework for reasoning about the impact of descriptions.} Upon first inspection, there appears to be a contradiction embedded in our results: we established that respondents care about information disclosures relevant to DP and are (in some cases) more willing to share their information when they are assured that these information disclosures will not occur. But, we found that offering respondents DP did \emph{not} increase their willingness to share information, no matter the description. At first glance, these results might seem to indicate that respondents did not understand the descriptions at all. However, the results in \cref{tab:betterbydefn} show that respondents had higher privacy expectations when presented with some descriptions. One would expect that these higher expectations would lead to higher willingness to share, in line with our first results.

To resolve this tension, we recall that not every respondent cared about every kind of information disclosure (\cref{sec:resultsbranch34}). While many respondents cared about each kind of disclosure, none of the information disclosures were important to more than 60\% of respondents. Thus, there was almost certainly misalignment between the disclosures that mattered to a given respondent and the disclosures that were influenced by the DP description they were shown. For instance, imagine a potential user cared about the \emph{Share} expectation, but was presented with the \emph{Trust} description. This potential user's higher expectations for \emph{Hack} and \emph{Data analyst} disclosures would likely do little to raise their willingness to share.

These results suggest a framework for understanding how descriptions of DP influence a user's willingness to share information (visualized in \cref{fig:framework}). When users encounter a differentially private system, they already have privacy preferences and concerns. When a user sees a description of DP, the user's expectations about certain information flows may increase. If the ways in which their privacy expectations increase aligns with the types of information disclosure about which they are concerned, they may be more likely to share their private information.  A key takeaway from this framework is that a clear and concise description of DP may not be enough to raise user's willingness to share. Instead, it is important that a description speaks to users' concerns directly and be tailored to address those concerns, as we discuss below. 
%

\medskip\noindent
\textbf{Need for new descriptions.} It is very evident from our results that the way DP is described in-the-wild is insufficient to help users make informed decisions. There is no consistency or standardization in the language organizations use. Thus, characterizing the way users might see DP described required us to identify the six descriptive themes used in this work. The themes present in these descriptions seem to haphazardly raise users' expectations. This is especially concerning given the differences between local and central DP; if descriptions are not carefully tailored to the model, they may mislead users about the privacy properties of the system. Indeed, \cref{fig:distributions} shows that the existing descriptions of DP do little to correctly set expectations, no matter the deployment model.

We note that the simple descriptions that we showed respondents in our surveys are not completely ineffective or without use. For instance, using our \emph{Risk} description may be appropriate for a local DP deployment as it raised expectations broadly. However, because these descriptions do nothing to increase participation, they may not achieve the goals of system designers. 

There are two main alternatives for improving the state of DP descriptions. First, one could take the approach of Xiong et al. \cite{SP:XWLJ20}, carefully constructing descriptions of DP and training users to understand those descriptions. However, Xiong et. al.'s results indicate that such an approach is difficult: a significant number of users were unable to correctly answer test questions about DP after viewing carefully crafted descriptions. An alternative approach, which prior work on privacy beyond DP suggests may be particularly effective~\cite{schaub2017designing}, is to explicitly inform users about the risks posed to their information. For instance, a description of a central DP system might specify that information will not be leaked through any graphs or informational chats, but could still be leaked to the other entities listed above. This would be similar to the privacy nutrition labels proposed by Kelley et al. \cite{kelley2009nutrition}. Such descriptions of DP could allow users to make an informed information sharing decision without requiring them to build a comprehensive mental model of the technical details of DP techniques. That said, technical details and parameter choices for DP deployment (e.g., the value of $\epsilon$) have important implications for user privacy and, as such, future work should also explore how best to communicate these technical nuances in meaningful ways.
\section{Conclusion}
In this work we studied DP from the user's perspective, focusing on how users' privacy expectations relate to DP as they are likely to encounter it in-the-wild. We showed that the privacy concerns about which users care can be addressed by DP, but the varied ways in which DP is described set user expectations in a haphazard, and often inaccurate, manner. Our results indicate that the interaction between user's intrinsic privacy concerns and the ways in which descriptions of DP set user expectations informs a user's willingness to share their information under differentially private guarantees. Our work posits a novel framework for understanding this interplay and suggests concrete directions for developing better descriptions of DP that directly and accurately address user privacy concerns. 
\section{Acknowledgments}
The first author was supported in part by The Defense Advanced Research Projects Agency (grant number W911NF-21-1-0371), NSF
grants CNS-1850187 and CNS-1942772 (CAREER), a
Mozilla Research Grant, and a JPMorgan Chase Faculty Research Award. Part of this work was completed while the first author was at Georgia Institute of Technology. The second author is supported by the National Science Foundation under Grant \#2030859 to the Computing Research Association for the CIFellows Project and The Defense Advanced Research Projects Agency under Agreement No. HR00112020021.  Part of this work was completed while the second author was at Johns Hopkins University. Part of this work was completed while the third author was at Microsoft Research. Any opinions, findings and conclusions or recommendations expressed in this material are those of the author(s) and do not necessarily reflect the views of the United States Government or DARPA. 


\balance
\bibliographystyle{plain}

\appendix
\section{Complete Survey Descriptions}\label{apx:surveys}
For completeness, we include the language and flow of our two surveys below.  Demographics for the two samples are included in \cref{tab:surveydemographics}.

\begin{table*}[!ht]
\centering%
\small

\begin{tabularx}{\linewidth}{l|XXX|XXX}
\toprule
&\multicolumn{3}{c|}{\textbf{Survey One} (n=1,216)} & \multicolumn{3}{c}{\textbf{Survey Two} (n=1,208)} \\
\midrule
 & \textbf{Percent} & \textbf{Mean} & \textbf{Stdev} & \textbf{Percent} & \textbf{Mean} & \textbf{Stdev} \\ 
\midrule
\midrule

Age & - & 37.09 & 12.00 & - & 37.39 & 11.16 \\
\midrule
Woman & 42.92\% & - & - & 40.64\% & - & - \\
Man & 56.08\% & - & - & 58.36\% & - & - \\
\midrule
Black  & 12.00\% & - & - & 11.92\% & - & -\\
White & 76.15\% & - & - & 76.98\% & - & - \\
Hispanic  & 7.48\% & - & - & 7.12\% & - & - \\
Asian & 9.12\% & - & - & 7.53\% & - & - \\
Native American & 1.64\% & - & - & 1.32\% & - & - \\
\midrule
Edu. High school or less & 9.04\% & - & - & 9.10\% & - & - \\
Edu. Some College & 23.60\% & - & - & 22.93\% & - & - \\
Edu. Bachelor's or above & 66.78\% & - & - & 67.54\% & - & - \\
Income & - & US\$61.6K & US\$42.9K & - & US\$61.5K & US\$41.3K \\
Internet Skills (1-5)  & - & 2.19 & .89 & - & 2.28 & .86 \\

\bottomrule
\end{tabularx}
\caption{\label{tab:surveydemographics} Survey demographics.}
\end{table*}

\subsection{Complete Description of Survey One}\label{apx:surveyone}
Respondents were randomized into either the salary scenario or the medical scenario, described below. 

\medskip \noindent
\textbf{Salary Scenario:} Imagine that you work in the banking industry. You are friends with a group of other people who work in banking companies in your city. One of your friends is part of a transparency initiative that is trying to publish general statistics about pay in the banking industry. As part of this initiative, they have asked everyone in the group to share their salaries and job titles using an online web form on the initiative’s website.

\medskip \noindent
\textbf{Medical Scenario:} Imagine that during your next doctor’s visit, your primary care doctor informs you that they are part of a non-profit organization trying to push the boundaries of medical research. This non-profit is asking patients around the country to share their medical records, which will be used to help medical research on improving treatment options and patient care. Your doctor, with your permission, can facilitate the non-profit getting the information they need. 

\medskip \noindent
\textbf{Questions:} Answer options for each questions presented in $\langle\rangle$.  Text differences between the two scenarios presented in italics inside brackets.
\begin{itemize}[--,leftmargin=*]
    \item Which of the following would you want to know \underline{before} deciding whether or not to share your \emph{[salary/medical history]}?  Select as many as apply.
    \begin{itemize}
        \item $\langle$\emph{[Whether your friend could learn your salary/Whether other doctors involved in the non-profit could learn your medical history]}
        \item Whether a criminal or foreign government could steal your \emph{[salary/medical history]}
        \item Whether law enforcement could accesses your \emph{[salary/medical history]} by obtaining a warrant
        \item Whether data analyst at the initiative could see your \\ \emph{[salary/medical history]}
        \item Whether graphs and charts created by the initiative could reveal your \emph{[salary/medical history]}
        \item Whether your \emph{[salary/medical history]} could be shared with another organization
        \item Other [free response]
    \end{itemize}
\end{itemize}

For each of the non-other options selected by the respondent, they were shown one of the follow three options, selected independently at random.
\begin{itemize}[--,leftmargin=*]
    \item \emph{[leak entity]} will not learn your \emph{[salary/medical history]}.
    \item \emph{[leak entity]} might learn your \emph{[salary/medical history]}.  The chance this happens is about the same as the chance that your bank account will be compromised (accessed by a person who you did not intend to gain access to) as part of a data breach in the next year.
    \item \emph{[leak entity]} might learn your \emph{[salary/medical history]}. The chance this happens is higher than the chance that your bank account will be compromised (accessed by a person who you did not intend to gain access to) as part of a breach in the next year.
\end{itemize}

Finally, respondents were asked:
\begin{itemize}[--,leftmargin=*]
    \item Would you be willing to share your \emph{[salary/medical record]} with the \emph{[initiative/non-profit]}? 
    
    $\langle$Yes, No, I'm not sure, Prefer not to answer$\rangle$
    \begin{itemize}[--]
        \item \emph{[If Yes]} Why would you be willing to share your \emph{[salary/medical record]}?
        \item \emph{[If No]} Why would you not be willing to share your \\ \emph{[salary/medical record]}?
        \item \emph{[If I'm not sure]} Why are you unsure whether you would be willing to share your \emph{[salary/medical record]} with the \emph{[initiative/non-profit]}?
    \end{itemize}
\end{itemize}

\subsection{Complete Description of Survey Two}\label{apx:surveytwo}
Respondents were randomized into either the salary scenario or the medical scenario, described below.  In both scenarios, respondents were randomly shown a description of differential privacy, listed after the scenarios.

\medskip \noindent
\textbf{Salary Scenario:} Imagine that you work in the banking industry. You are friends with a group of other people who work in banking companies in your city. One of your friends is part of a transparency initiative that is trying to publish general statistics about pay in the banking industry. As part of this initiative, they have asked everyone in the group to share their salaries and job titles using an online web form on the initiative’s website. \textit{[description from the list of differential privacy descriptions, shown below.]} 

In this survey we are going to ask you a series of questions about a hypothetical scenario. Please do your best to imagine yourself in this scenario and answer the questions as if you were actually making the decisions about which you will be asked.

\medskip \noindent
\textbf{Medical Scenario:} Imagine that during your next doctor’s visit, your primary care doctor informs you that they are part of a non-profit organization trying to push the boundaries of medical research. This non-profit is asking patients around the country to share their medical records, which will be used to help medical research on improving treatment options and patient care. Your doctor, with your permission, can facilitate the non-profit getting the information they need. \textit{[description from the list of differential privacy descriptions, shown below.]} 

In this survey we are going to ask you a series of questions about a hypothetical scenario. Please do your best to imagine yourself in this scenario and answer the questions as if you were actually making the decisions about which you will be asked.

\medskip \noindent
\textbf{List of Differential Privacy Descriptions:} Names of the description, shown in italics inside parenthesis, were not show to respondents.
\begin{itemize}[--,leftmargin=*]
    \item \emph{(Control:) no additional text} 
    \item \emph{(Unsubstantial:)} To reduce the intrusion into personal privacy, your friend says they will use a technique called differential privacy. Differential privacy is the gold standard in data privacy and protection and is widely recognized as the strongest guarantee of privacy available.
    \item \emph{(Techniques:)} To reduce the intrusion into personal privacy, your friend says they will use a technique called differential privacy. Differential Privacy injects statistical noise into collected data in a way that protects privacy without significantly changing conclusions. 
    \item \emph{(Enables:)} To reduce the intrusion into personal privacy, your friend says they will use a technique called differential privacy. Differential Privacy allows analysts to learn useful information from large amounts of data without compromising an individual's privacy.
    \item \emph{(Trust:)} To reduce the intrusion into personal privacy, your friend says they will use a technique called differential privacy. Differential privacy is a novel, mathematical technique to preserve privacy which is used by companies like Apple and Uber.
    \item \emph{(Risk:)} To reduce the intrusion into personal privacy, your friend says they will use a technique called differential privacy. Differential privacy protects a user’s identity and the specifics of their data, meaning individuals incur almost no risk by joining the dataset.
    \item \emph{(Technical:)} To reduce the intrusion into personal privacy, your friend says they will use a technique called differential privacy. Differential privacy ensures that the removal or addition of a single database item does not (substantially) affect the outcome of any analysis. It follows that no risk is incurred by joining the database, providing a mathematically rigorous means of coping with the fact that distributional information may be disclosive.
\end{itemize}

\medskip \noindent
\textbf{Questions:} Answer options for each questions presented in $\langle\rangle$.  Text differences between the two scenarios presented in italics inside brackets.  The names of the data leaks used in the main body of the text are shown in italics inside parenthesis and were not shown to users.
\begin{itemize}[--,leftmargin=*]
    \item Would you be willing to share your \emph{[salary/medical record]} with the \emph{[initiative/non-profit]}? 
    
    $\langle$Yes, No, I'm not sure, Prefer not to answer$\rangle$
    \begin{itemize}[--]
        \item \emph{[If Yes]} Why would you be willing to share your \emph{[salary/medical record]}?
        \item \emph{[If No]} Why would you not be willing to share your \\ \emph{[salary/medical record]}?
        \item \emph{[If I'm not sure]} Why are you unsure whether you would be willing to share your \emph{[salary/medical record]} with the \emph{[initiative/non-profit]}?
    \end{itemize}
\end{itemize}

\medskip \noindent
For each of the following statements, please indicate if you expect the following to be true or false if you share your salary and job title as part of this initiative.
\begin{itemize}[--,leftmargin=*]
    \item \emph{(Organization:)} \emph{[My friend will not be able to learn my salary and job title/The contents of my medical record will be stored only by my doctor's office, and will not be stored by the non-profit]} 
    
    $\langle$ Yes, No, I don't know $\rangle$
    
    \item \emph{(Hack:)} A criminal or foreign government that hacks the \emph{[transparency initiative/non-profit]} could learn my \emph{[salary and job title/medical history]}
    
    $\langle$Yes, No, I don't know$\rangle$
    
    \item \emph{(Law Enforcement:)} A law enforcement organization could access my \emph{[salary and job title/medical history]} with a court order requesting this data from the \emph{[transparency initiative/non-profit]}
    
    $\langle$Yes, No, I don't know$\rangle$

    \item \emph{(Data Analyst:)} A data analyst working \emph{[on/for]} the \emph{[salary transparency initiative/non-profit]} could learn my exact \emph{[salary and job title/medical history]}
        
    $\langle$Yes, No, I don't know$\rangle$
    
    \item \emph{(Graphs:)} Graphs or informational charts created using information given to the \emph{[salary transparency initiative/non-profit]} could reveal my \emph{[salary and job title/medical history]}
        
    $\langle$Yes, No, I don't know$\rangle$
    
    \item \emph{(Share:)} Data that the \emph{[salary transparency initiative/non-profit]} shares with other organizations doing \emph{[salary/medical]} research could reveal my \emph{[salary and job title/medical history]}
        
    $\langle$Yes, No, I don't know$\rangle$
    
\end{itemize}

\end{document}